\documentclass[notitlepage,nofootinbib,preprintnumbers,amssymb,superscriptaddress]{revtex4-2}
\usepackage{amsfonts,amssymb,mathtools,graphicx,color,bm,orcidlink}
\definecolor{ultramarine}{rgb}{0.07, 0.04, 0.56}
\definecolor{cadmiumgreen}{rgb}{0.0, 0.42, 0.24}
\definecolor{indigo(dye)}{rgb}{0.0, 0.25, 0.42}
\usepackage{hyperref}
\hypersetup{
colorlinks=true,
citecolor=ultramarine,
linkcolor=cadmiumgreen,
urlcolor=indigo(dye),
}

\let\originalleft\left
\let\originalright\right
\renewcommand{\left}{\mathopen{}\mathclose\bgroup\originalleft}
\renewcommand{\right}{\aftergroup\egroup\originalright}
\usepackage{autobreak}

\newcommand{\fr}[2]{\frac{#1}{#2}}
\newcommand{\pa}{\partial}
\newcommand{\ti}{\tilde}
\newcommand{\na}{\nabla}
\newcommand{\bra}[1]{\left( #1 \right)}  
\newcommand{\brb}[1]{\left[ #1 \right]}  
  
\newcommand{\be}{\begin{equation}}  
\newcommand{\ee}{\end{equation}}
\newcommand{\bem}{\begin{bmatrix}}
\newcommand{\eem}{\end{bmatrix}}

\newcommand{\ga}{g_{a\gamma\gamma}}

\newcommand{\mn}{{\mu \nu}}

\allowdisplaybreaks

\begin{document}

\preprint{RESCEU-10/25}

\title{Resonance of black hole quasinormal modes in coupled systems}

\author{Takuya Takahashi\,\orcidlink{0000-0002-4894-6108}}
\email{takuya.takahashi@resceu.s.u-tokyo.ac.jp}
\affiliation{Research Center for the Early Universe (RESCEU), Graduate School of Science, The University of Tokyo, Tokyo 113-0033, Japan}

\author{Hayato Motohashi\,\orcidlink{0000-0002-4330-7024}}
\affiliation{Department of Physics, Tokyo Metropolitan University, 1-1 Minami-Osawa, Hachioji, Tokyo 192-0397, Japan}

\author{Kazufumi Takahashi\,\orcidlink{0000-0002-4070-1675}}
\affiliation{Department of Physics, College of Humanities and Sciences, Nihon University, Tokyo 156-8550, Japan}
\affiliation{Center for Gravitational Physics and Quantum Information, Yukawa Institute for Theoretical Physics, Kyoto University, 606-8502, Kyoto, Japan}

\begin{abstract}
Black hole quasinormal modes (QNMs) can exhibit resonant excitations associated with avoided crossings in their complex frequency spectrum.
Such resonance phenomena can serve as novel signatures for probing new physics, where additional degrees of freedom are commonly introduced.
Motivated by this possibility, we investigate QNMs in systems where multiple degrees of freedom are coupled with each other, and introduce a definition of excitation factors suitable for such systems.
To demonstrate our formulation, we apply it to a black hole in the Einstein-Maxwell-axion theory, where we find that avoided crossings can appear even between longest-lived modes originating from the fundamental modes of different degrees of freedom,
in contrast to the Kerr case in General Relativity.
We show that the excitation factors are indeed amplified as a manifestation of resonance at parameter values corresponding to the avoided crossings.
\end{abstract}

\maketitle

\section{Introduction}\label{sec:intro}

The advancement of gravitational wave observations has opened up new opportunities to explore physics beyond known frameworks, including modified gravity and new particles.
To date, numerous gravitational wave events from binary mergers of compact objects have been detected~\cite{LIGOScientific:2016aoc,KAGRA:2021vkt}.
The gravitational waves emitted as the black hole (BH) formed immediately after the merger settles into a stationary state are known as ``ringdown", and are well described by BH perturbation theory~\cite{Vishveshwara:1970zz,Press:1971wr,Teukolsky:1973ha}.
The waveform is characterized by quasinormal modes (QNMs), which reflect the properties of the BH spacetime~\cite{Chandrasekhar:1975zza,Kokkotas:1999bd,Dreyer:2003bv,Berti:2005ys,Berti:2009kk,Konoplya:2011qq,Baibhav:2023clw} as well as underlying theories of gravity (see, {\it e.g.}, Refs.~\cite{Cardoso:2019mqo,McManus:2019ulj,Mukohyama:2023xyf}).
Although the BH perturbation theory has a long-standing history, certain theoretical aspects remain not fully understood, and notable progress has been made in recent years.

BH QNMs possess an infinite set of complex eigenfrequencies, 
where the mode with the smallest damping rate is called the fundamental mode and those with larger damping rates are called overtones.
In General Relativity in vacuum, that is, in the case of Kerr BHs~\cite{Kerr:1963ud}, it has long been known that 
certain overtones exhibit 
a peculiar behavior in the complex plane~\cite{Onozawa:1996ux,Berti:2003jh,Berti:2004md,Cook:2014cta}.
Recently, Ref.~\cite{Motohashi:2024fwt} proposed that this behavior arises from avoided crossing 
between overtone modes.
Moreover, it was reported that, as the eigenfrequencies of the two modes approach each other, excitation factors~\cite{Leaver:1986gd,Berti:2006wq}, which characterize how easily each QNM is excited, undergo a significant amplification.
This phenomenon can be interpreted as a resonance between different overtones and is expected to serve as a distinctive signature for probing the properties of the BH spacetime.

However, in the case of Kerr BHs, the resonance occurs between overtones with relatively high damping rates.
As a result, extracting the resonance structure from actual waveforms remains an open issue in current gravitational wave research, particularly in connection with data analysis~\cite{Giesler:2019uxc,Cook:2020otn,Oshita:2021iyn,Ma:2022wpv,Baibhav:2023clw,Takahashi:2023tkb,Cheung:2023vki,Oshita:2024wgt,Mitman:2025hgy,Oshita:2025ibu,Yang:2025dbn}.
Moreover, the resonance arises only at high values of the BH spin parameter, which may present additional observational challenges.

Having said that, there are also other possibilities where we could see such a resonance.
Indeed, the resonance also occurs for electromagnetic and scalar perturbations around the Kerr BH~\cite{Motohashi:2024fwt,Lo:2025njp}.
More generally, it universally occurs when two QNMs approach each other. 
Hence, the resonant excitation of QNMs has the potential to serve as a probe for uncovering new physics~\cite{Motohashi:2024fwt}.
For example, in modified gravity theories such as scalar-tensor theories, additional degree(s) of freedom are often introduced on top of the metric.
Similarly, hypothetical fields that could account for, {\it e.g.}, dark matter may interact through gravity.
In either case, such extra degrees of freedom generally emerge in frameworks beyond General Relativity and/or the Standard Model.
Motivated by these considerations, in this work we investigate the resonance of QNMs in BH systems with multiple degrees of freedom.
In particular, we provide a general formulation of excitation factors for the case where these degrees of freedom are coupled at the level of the perturbation equations.

As a concrete example to demonstrate this formulation, we apply it to a BH solution in the Einstein-Maxwell-axion (EMA) theory.
In the absence of an axion field, a static and spherically symmetric BH solution in the Einstein-Maxwell theory is known as the Reissner-Nordstr\"{o}m BH, which carries electric (and possibly magnetic) charge~\cite{Reissner:1916cle}.
In this case, the degrees of freedom originating from gravitational and electromagnetic fields are decoupled for an appropriate choice of master variables, and each perturbation can be described by a single master equation~\cite{Moncrief:1974gw,Moncrief:1974ng,Moncrief:1975sb}.
However, the situation changes in the presence of an axion field.
When the BH carries only electric charge, the axion admits only a trivial configuration.
If the BH carries both electric and magnetic charges, on the other hand, a nontrivial axion configuration is allowed as a secondary hair~\cite{Lee:1991jw,Filippini:2019cqk,DeFelice:2024eoj}.
Interestingly, even when the background BH carries only electric charge and does not have axion hair, the degrees of freedom associated with the odd-parity modes of the gravitational, electromagnetic, and axion fields are coupled at the level of the perturbation~\cite{Boskovic:2018lkj}.
As a result, the QNMs deviate from those of the Einstein-Maxwell theory, and resonant behavior may emerge due to the axion-photon coupling.

In this work, we compute the QNM frequencies in the EMA system, including higher overtones, using the continued fraction method.
By evaluating the excitation factors defined in our formalism, we demonstrate that the resonances accompanied by a significant enhancement of excitation factors appear in this system.
Unlike the resonances between overtones in the Kerr case in General Relativity, the resonances in the EMA system exhibit a qualitatively different behavior: the resonance takes place between longest-lived modes originating from the fundamental modes of different degrees of freedom. 
Moreover, to further understand the phenomenon, we 
discuss key features of the resonant excitation at avoided crossing, following Ref.~\cite{Motohashi:2024fwt}.

The rest of this paper is organized as follows.
In Sec.~\ref{sec:formulation}, we provide a brief overview of BH QNMs and introduce the concept of excitation factors in the BH perturbation equations with coupled multiple degrees of freedom. 
In Sec.~\ref{sec:application}, we apply our formulation to the BH in the EMA theory.
In particular, we show numerical results for the QNM frequency spectra and their associated excitation factors, and examine the resonance behavior implied by them.
Finally, we draw our conclusions in Sec.~\ref{sec:conc}.
Throughout this paper, we use the units with $c=G=4\pi\epsilon_0=1$, with $c$ the speed of light in a vacuum, $G$ the gravitational constant, and $\epsilon_0$ the permittivity in a vacuum.


\section{Formulation: Quasinormal modes and excitation factors in coupled systems}
\label{sec:formulation}
In this section, we briefly review the QNMs of BHs and introduce the excitation factors when multiple degrees of freedom are coupled in the BH perturbation equations.

\subsection{Quasinormal modes}
\label{subsec:formulation_QNM}

Let us consider $N$ perturbation variables, denoted as $\vec{\Psi}(t,r)=(\Psi_1,\Psi_2,\cdots,\Psi_N)^{\rm t}$, governed by the following equation:
\begin{align}\label{eq:Peq_t}
    \left(-\frac{\pa^2}{\pa t^2}+\frac{\pa^2}{\pa r_\ast^2}\right)\vec{\Psi}(t,r)-\bm{V}(r)\vec{\Psi}(t,r)=0~,
\end{align}
where $\bm{V}(r)$ is an $N\times N$ potential matrix, and $r_\ast$ is the tortoise coordinate.
Note that equations of motion of perturbations on a BH background are typically written in this form.
We assume $r_\ast\to-\infty$ as $r\to r_{\rm h}$ and $r_\ast\to\infty$ as $r\to\infty$, where $r_{\rm h}$ denotes the location of the BH event horizon.\footnote{Here and in what follows, for simplicity, we assume that the $N$ degrees of freedom share the same horizon radius.
Note that this is the case for the EMA system that we will study in Sec.~\ref{sec:application}.
In general, different degrees of freedom can have different horizon radii, implying that one field can probe the BH interior of another.
This offers an interesting possibility of energy extraction from the BH and a characteristic late-time relaxation~\cite{Cardoso:2024qie}.}
In terms of the Laplace transform,
\begin{align}\label{eq:Laplace}
    \vec{\Psi}(t,r)=\frac{1}{2\pi}\int_{-\infty+ic}^{\infty+ic}d\omega\ \vec{\hat{\Psi}}(\omega,r)e^{-i\omega t}~,
\end{align}
with $c>0$, Eq.~\eqref{eq:Peq_t} can be written as
\begin{align}\label{eq:Peq_f}
    \left(\frac{d^2}{dr_\ast^2}+\omega^2\right)\vec{\hat{\Psi}}(\omega,r)-\bm{V}(r)\vec{\hat{\Psi}}(\omega,r)=0~.
\end{align}
Here,  we assume $\bm{V}(r)\to0$ for $r\to r_{\rm h}$ and $r\to\infty$ for simplicity, which corresponds to considering massless fields.

QNMs for coupled systems are given by particular solutions of Eq.~\eqref{eq:Peq_f} that satisfy physically appropriate boundary conditions; purely in-going wave at the horizon and purely out-going wave at infinity~\cite{Pani:2013pma,Chaverra:2016ttw}. 
We define the ``in-mode" functions as the homogeneous solutions which consist of purely in-going waves at the horizon;
\begin{align}\label{eq:inmode}
    \vec{\Psi}^{({\rm in},\alpha)}\to e^{-i\omega r_\ast}\vec{\delta}_\alpha~, \quad r_\ast\to-\infty~,
\end{align}
where $\vec{\delta}_\alpha=(\delta_{\alpha1},\delta_{\alpha2},\cdots,\delta_{\alpha N})^{\rm t}$, and $\delta_{\alpha\beta}$ is the Kronecker delta.
Also, we define the ``up-mode" functions as the homogeneous solutions which consist of purely out-going waves at infinity;
\begin{align}\label{eq:upmode}
    \vec{\Psi}^{({\rm up},\alpha)}\to e^{+i\omega r_\ast}\vec{\delta}_\alpha~, \quad r_\ast\to+\infty~.
\end{align}
Now, the Wronskian matrix of the above solutions is defined as
\begin{align}
    &\bm{W}\left[\vec{\Psi}^{({\rm in},1)},\cdots,\vec{\Psi}^{({\rm in},N)},\vec{\Psi}^{({\rm up},1)},\cdots,\vec{\Psi}^{({\rm up},N)}\right] \notag \\
    &=\begin{pmatrix}
        \vec{\Psi}^{({\rm in},1)}&\cdots&\vec{\Psi}^{({\rm in},N)} & \vec{\Psi}^{({\rm up},1)}&\cdots&\vec{\Psi}^{({\rm up},N)} \\
        \partial_{r_\ast}\vec{\Psi}^{({\rm in},1)}&\cdots& \partial_{r_\ast}\vec{\Psi}^{({\rm in},N)} &  \partial_{r_\ast}\vec{\Psi}^{({\rm up},1)}&\cdots& \partial_{r_\ast}\vec{\Psi}^{({\rm up},N)}
    \end{pmatrix}~, \label{eq:Wdef}
\end{align}
which is a $2N\times 2N$ matrix.
In the following, we simply denote the Wronskian for the set of solutions of Eq.~\eqref{eq:Wdef} as $\bm{W}$.
The determinant of the Wronskian is a function of $\omega$ and does not depend on $r_\ast$.
QNM frequencies are determined by those values of $\omega$ that satisfy
\begin{align}
    \det \bm{W}(\omega)=0~.
\end{align}
For such frequencies, there exist constant complex vectors~$\vec{a}$ and $\vec{b}$ such that
\begin{align}
    \vec{\hat{\Psi}}(\omega,r)=\left(\vec{\Psi}^{({\rm in},1)},\cdots,\vec{\Psi}^{({\rm in},N)}\right)\vec{a}=\left(\vec{\Psi}^{({\rm up},1)},\cdots,\vec{\Psi}^{({\rm up},N)}\right)\vec{b}~,
\end{align}
where $e^{-i\omega t}\vec{\hat{\Psi}}(\omega,t)$ corresponds to the QNM.
For ${\rm Im}[\omega]<0$, it represents a damping wave in time.
Due to the boundary conditions at two points and dissipation, the QNM frequencies are discrete complex eigenvalues.
One usually arranges the QNM frequencies in order of increasing $|{\rm Im}[\omega]|$ and labels them with $n=0,1,2,\cdots$.
The $n=0$ mode, {\it i.e.}, the longest-lived mode, is called the fundamental mode, while the other modes are referred to as higher overtone modes.

Let us comment on the labeling of overtones in coupled systems.
Suppose that there exists a decoupling limit where the $N$ degrees of freedom are decoupled.
In this limit, the QNM spectrum reduces to a union of QNM spectra for the $N$ decoupled degrees of freedom.
Therefore, one expects that each QNM of the coupled system can be traced back to a QNM of one of the degrees of freedom in the decoupling limit.\footnote{Of course, it may in principle be possible that some QNMs (dis)appear at some critical values of the coupling. Nevertheless, in Sec.~\ref{sec:application}, we mainly focus on modes which have the decoupling limit.}
For such QNMs, it is sometimes useful to classify them in terms of the corresponding QNMs in the decoupling limit, as we will do in Sec.~\ref{sec:application}.
Note in particular that what is the fundamental mode of some degree of freedom in the decoupling limit may be regarded as an overtone mode in the QNM spectrum of the coupled system when there exist other longer-lived QNM(s) originating from other degrees of freedom.

\subsection{Excitation factors}
\label{subsec:formulation_EF}

\subsubsection{Waveform}

Next, we introduce excitation factors for coupled systems, which characterize how easily each QNM is excited
in response to an external source for each QNM. 
For this purpose, we consider Eq.~\eqref{eq:Peq_f} with an added source term,
\begin{align}
    \left(\frac{d^2}{dr_\ast^2}+\omega^2\right)\vec{\hat{\Psi}}(\omega,r)-\bm{V}(r)\vec{\hat{\Psi}}(\omega,r)=\vec{S}(r)~.
\end{align}
The solutions of these equations, which satisfy the boundary conditions of Eqs.~\eqref{eq:inmode} and \eqref{eq:upmode}, can be constructed using the Green's matrix method~\cite{Reid:1931,Sisman:2009mk,OuldElHadj:2024psw}
(see also App.~\ref{App:Green}).
Formally, such a solution is expressed as
\begin{align}\label{eq:sol}
    \vec{\hat{\Psi}}(\omega,r_\ast)=\int dr_\ast\ \bm{G}(r_\ast,r_\ast')\vec{S}(r_\ast')~,
\end{align}
where the Green's matrix is given by
\begin{align}\label{eq:Green}
\bm{G}(r_\ast,r_\ast')=\begin{cases}
-\bm{U}^{\rm L}_{N\times 2N}\bm{W}^{({\rm in})}(r_\ast)\bm{W}^{-1}(r_\ast')\bm{L}^{\rm D}_{2N\times N}~, \quad &r_\ast<r_\ast'~, \\
\bm{U}^{\rm L}_{N\times 2N}\bm{W}^{({\rm up})}(r_\ast)\bm{W}^{-1}(r_\ast')\bm{L}^{\rm D}_{2N\times N}~, \quad &r_\ast>r_\ast'~.
\end{cases}
\end{align}
Here, $\bm{W}^{({\rm in})}$ and $\bm{W}^{({\rm up})}$ are $2N\times 2N$ matrices constructed from the Wronskian matrix~\eqref{eq:Wdef} as
\begin{align}
\bm{W}^{({\rm in})}=\bm{W}\begin{pmatrix}
\bm{I} & \bm{O} \\
\bm{O} & \bm{O}
\end{pmatrix}~, \quad
\bm{W}^{({\rm up})}=\bm{W}\begin{pmatrix}
\bm{O} & \bm{O} \\
\bm{O} & \bm{I}
\end{pmatrix}~,
\end{align}
where $\bm{O}$ and $\bm{I}$ denote the $N\times N$ zero matrix and identity matrix, respectively.
In addition, selection matrices are defined as
\begin{align}
\bm{U}^{\rm L}_{N\times 2N}=\begin{pmatrix} \bm{I} & \bm{O} \end{pmatrix}~, \quad \bm{L}^{\rm D}_{2N\times N}=\begin{pmatrix} \bm{O} \\ \bm{I} \end{pmatrix}~.
\end{align}
Focusing on the solution far away from the source, we have
\begin{align}
\bm{U}^{\rm L}_{N\times 2N}\bm{W}^{({\rm up})}= e^{i\omega r_\ast}\begin{pmatrix} \bm{O} & \bm{I} \end{pmatrix}\equiv e^{i\omega r_\ast}\bm{U}^{\rm R}_{N\times 2N}~.
\end{align}
Then, Eq.~\eqref{eq:sol} can be written as
\begin{align}\label{eq:sol2}
\vec{\hat{\Psi}}(\omega,r_\ast)=\frac{e^{i\omega r_\ast}}{\det\bm{W}}\int dr_\ast'\ \bm{U}^{\rm R}_{N\times 2N}\tilde{\bm{W}}(r_\ast')\bm{L}^{\rm D}_{2N\times N}\vec{S}(r_\ast')~,
\end{align}
where we have used $\bm{W}^{-1}=(\det \bm{W})^{-1}\tilde{\bm{W}}$, and $\tilde{\bm{W}}$ is the adjugate matrix of $\bm{W}$.

Now, we define the amplitude of the in-going wave and the out-going wave from the asymptotic behavior of the in-mode functions as
\begin{align}
\vec{\Psi}^{({\rm in},\alpha)}\to
\begin{cases}
e^{-i\omega r_\ast}\vec{\delta}_\alpha~, &r_\ast\to-\infty~, \\
e^{-i\omega r_\ast}\vec{A}^{({\rm in},\alpha)}+e^{+i\omega r_\ast}\vec{A}^{({\rm out},\alpha)}~, \quad &r_\ast\to+\infty~.
\end{cases}
\end{align}
This allows $\det\bm{W}$ to be written as
\begin{align}
    \det\bm{W}=(2i\omega)^N\det\bm{A}^{(\rm in)}~,
\end{align}
where
\begin{align}
    \bm{A}^{(\rm in)}=\left(\vec{A}^{({\rm in},1)},\vec{A}^{({\rm in},2)},\cdots,\vec{A}^{({\rm in},N)}\right)~.
\end{align}
From the above, by performing the inverse Laplace transform~\eqref{eq:Laplace} on Eq.~\eqref{eq:sol2}, we obtain the expression for the waveform in terms of the time domain.
Extracting only contributions from poles in the contour corresponding to the QNMs, the retarded wave is given as
\begin{align}\label{eq:Psi_t}
\vec{\Psi}(t,r)=\sum_{\alpha,n} \frac{i}{(2i\omega_{\alpha n})^N}\left.\left(\frac{d}{d\omega}\det\bm{A}^{({\rm in})}\right)^{-1}\right|_{\omega=\omega_{\alpha n}}e^{-i\omega_{\alpha n}(t-r_\ast)}\int dr_\ast'\ \bm{U}^{\rm R}_{N\times 2N}\tilde{\bm{W}}(r_\ast')\bm{L}_{2N\times N}\vec{S}(r_\ast')~,
\end{align}
where we have assumed for simplicity that each QNM has the decoupling limit and assigned to each QNM the labels corresponding to the associated perturbation variable~$\alpha$ and overtone number~$n$.
Here, the factor in front of the source integral tells us the information on the ease of excitation of the QNMs, which is essentially nothing but what is called the excitation factor.
However, this is still not what we want as there is an ambiguity associated with the overall normalization of the wavefunction.
In what follows, we discuss how to define the excitation factors without this ambiguity.

\subsubsection{Definition}

With the setup above, let us define the excitation factors in coupled systems.
In the case of a single degree of freedom, {\it i.e.}, $N=1$, the definition of the excitation factor is given by~\cite{Leaver:1986gd,Berti:2006wq}
\begin{align}\label{eq:EF1}
B_n^{\rm single}=\frac{A^{({\rm out})}}{2\omega_n}\left.\left(\frac{d}{d\omega}A^{({\rm in})}\right)^{-1}\right|_{\omega=\omega_n}~.
\end{align}
In linear perturbation theory, there is an ambiguity in the overall normalization of the wavefunction.
To avoid this ambiguity, Eq.~\eqref{eq:EF1} is normalized by the coefficient $A^{({\rm out})}$ of the out-going mode at infinity.

Now, we extend the above definition of excitation factors to the case where multiple degrees of freedom are coupled.
As in the case of single degree of freedom, we must normalize the prefactor in Eq.~\eqref{eq:Psi_t} so that the excitation factors are not affected by the overall rescaling of the wavefunction.
Also, we should adopt a normalization that is independent of the choice of basis of variables.
Thus, we define the excitations factors in general coupled systems as
\begin{align}\label{eq:EF}
B_{\alpha n}=\frac{i\det\bm{A}^{({\rm out})}}{(2i\omega_{\alpha n})^N}\left.\left(\frac{d}{d\omega}\det\bm{A}^{({\rm in})}\right)^{-1}\right|_{\omega=\omega_{\alpha n}}~,
\end{align}
where $\bm{A}^{({\rm out})}=(\vec{A}^{({\rm out},1)},\cdots,\vec{A}^{({\rm out},N)})$.
In practice, $\det\bm{A}^{({\rm out})}$ can be computed through
\begin{align}
    \det\bm{A}^{({\rm out})}=-(2i\omega)^{-N}\det\bm{W}\left[\vec{\Psi}^{({\rm in},1)},\cdots,\vec{\Psi}^{({\rm in},N)},\vec{\Psi}^{({\rm down},1)},\cdots,\vec{\Psi}^{({\rm down},N)}\right]~,
\end{align}
where the ``down-mode" functions~$\vec{\Psi}^{({\rm down},\alpha)}$ are the homogeneous solutions which consist of purely in-going waves at infinity; $\vec{\Psi}^{({\rm up},\alpha)}\to e^{-i\omega r_\ast}\vec{\delta}_\alpha$ for $r_\ast\to\infty$.
It is obvious that this definition reproduces Eq.~\eqref{eq:EF1} in the case of $N=1$.

Before closing this section, we mention two remarks.
First, the tortoise coordinate is defined up to integration constant, and therefore one could shift $r_*$ by an arbitrary constant~$c_\ast$, {\it i.e.}, $r_\ast\to r_\ast+c_\ast$.
Thus, there is an ambiguity in the phase of $A^{({\rm in/out},\alpha)}$, same as in the case with single degree of freedom.
Accordingly, the definition~\eqref{eq:EF} has an arbitrariness by a factor of $e^{2Ni\omega_{\alpha n} c_\ast}$.
However, for the purpose of observing the resonance structure, the ambiguity is not important since $\omega_{\alpha n}$ remains almost a constant near the resonance. 
Therefore, as long as a consistent definition of the tortoise coordinate is used throughout the calculation of the excitation factors, this ambiguity can be safely ignored.

Second, suppose that there exists a decoupling limit where $\bm{V}(r)$ in Eq.~\eqref{eq:Peq_t} is diagonalized.
In this limit, one can apply the definition~\eqref{eq:EF1} for a single field to each of the $N$ decoupled fields.
Then, comparing it with the decoupling limit of Eq.~\eqref{eq:EF} for a field~$\Psi_\alpha$, one finds that the latter yields an extra factor coming from the contribution of the other fields~$\Psi_\beta$ with $\beta\ne \alpha$.
Written explicitly, the extra factor takes the form of $(2i\omega)^{-(N-1)}\prod_{\beta\neq \alpha}A_\beta^{({\rm out},\beta)}/A_\beta^{({\rm in},\beta)}$.
Here, we have used the fact that $\det\bm{A}^{({\rm in/out})}=\prod_{\alpha=1}^N A_\alpha^{({\rm in/out},\alpha)}$ in the decoupling limit, where $A_\alpha^{({\rm in/out},\alpha)}$ is the $\alpha$ component of $\vec{A}^{({\rm in/out},\alpha)}$.
One could define excitation factors in such a way that this extra factor does not show up in the decoupling limit, {\it e.g.},
    \begin{align}
    \tilde{B}_{\alpha n}=\frac{\det\tilde{\bm{A}}^{({\rm out},\alpha)}}{2\omega_{\alpha n}}\left.\left(\frac{d}{d\omega}\det\bm{A}^{({\rm in})}\right)^{-1}\right|_{\omega=\omega_{\alpha n}}~, \label{alt_defB}
    \end{align}
where
\begin{align}
    \det\tilde{\bm{A}}^{({\rm out},\alpha)}&=-(2i\omega)^{-N}\det\bm{W}\left[\vec{\Psi}^{({\rm in},1)},\cdots,\vec{\Psi}^{({\rm in},N)},\vec{\Psi}^{({\rm up},1)},\cdots,\vec{\Psi}^{({\rm down},\alpha)},\cdots,\vec{\Psi}^{({\rm up},N)}\right] \notag \\
    &\to A_\alpha^{({\rm out},\alpha)}\prod_{\beta\neq \alpha}A_\beta^{({\rm in},\beta)} \quad ({\rm decoupling\ limit})~.
\end{align}
However, unfortunately, this definition depends on the choice of basis of the fields, as opposed to $B_{\alpha n}$ defined in Eq.~\eqref{eq:EF}.
Note that this ambiguity exists even when there is a basis where the equations of motion are completely decoupled.\footnote{Moreover, for the system studied in Sec.~\ref{sec:application}, we have observed an unwanted growth of $\tilde{B}_{\alpha n}$ at large coupling for our choice of basis.}
Hence, instead of Eq.~\eqref{alt_defB}, we shall adopt the definition~\eqref{eq:EF} of excitation factors in what follows.


\section{Application: Einstein-Maxwell-axion system}\label{sec:application}

In this section, we apply the formulation introduced in the previous section to a specific theory.
As an example where multiple degrees of freedom couple and the resonance between QNMs is expected, we consider the Einstein-Maxwell-axion (EMA) system.

The action of the EMA system is given by
    \begin{align}
    S=\frac{1}{4\pi}\int d^4x\sqrt{-g}\brb{
    \fr{1}{4}R-\fr{1}{4}F_\mn F^\mn-\fr{1}{2}g^{\mn}\pa_\mu\phi\pa_\nu\phi
    -\fr{1}{4}\ga\phi F_\mn\ti{F}^\mn}~,
    \label{eq:action_EMA}
    \end{align}
where $R$ denotes the Ricci scalar, $F_\mn\coloneqq \na_\mu A_\nu-\na_\nu A_\mu$ is the field strength of the electromagnetic (EM)\footnote{The abbreviation~EM (electromagnetic) should not be confused with EMA (Einstein-Maxwell-axion).} field~$A_\mu$, and $\tilde{F}^\mn\coloneqq \varepsilon^{\mn\alpha\beta}F_{\alpha\beta}/2$ with the totally antisymmetric tensor~$\varepsilon_{\mn\lambda\sigma}$ defined so that $\varepsilon^{0123}=1/\sqrt{-g}$.
The axion field~$\phi$ is a pseudo-scalar, and hence the action~\eqref{eq:action_EMA} is invariant under a parity transformation.
The coupling constant~$\ga$ controls the coupling between the EM and axion fields.
One could introduce a potential term~$-V(\phi)$ in the action~\eqref{eq:action_EMA} (see, {\it e.g.}, Refs.~\cite{Fernandes:2019kmh,Burrage:2023zvk,Melis:2024kfr}), but we omit it in the following analysis for simplicity.
Here and in what follows, we adopt units where $c=G=4\pi\epsilon_0=1$, with $c$ the speed of light in a vacuum, $G$ the gravitational constant, and $\epsilon_0$ the permittivity in a vacuum.
Note also that we have normalized $\phi$ so that the ratio of the coefficients is rational.

\subsection{Black hole background}

The equations of motion for the EMA system are given by the following:
\begin{align}
&G_\mn=2T_\mn~, \label{eq:Eeq} \\
&\nabla_{\nu}\left(F^\mn+\ga \phi \tilde{F}^\mn\right)=0~, \label{eq:Meq} \\
&\nabla_\mu\nabla^\mu\phi=\frac{\ga}{4}F_\mn\tilde{F}^{\mu\nu}~. \label{eq:KGeq}
\end{align}
Here, $G_\mn$ is the Einstein tensor and the stress-energy tensor is given by
    \begin{align}
    T_{\mu\nu}=F_{\mu}^{\ \alpha}F_{\nu\alpha}-\frac{1}{4}g_{\mu\nu}F_{\alpha\beta}F^{\alpha\beta}+\partial_\mu\phi\partial_\nu\phi-\frac{1}{2}g_{\mu\nu}g^{\alpha\beta}\partial_\alpha\phi\partial_\beta\phi~.
    \end{align}
Note that the term with $g_{a\gamma\gamma}$ does not contribute to $T_{\mu\nu}$.

As a background solution, we consider a spherically symmetric and electrically charged BH.
When the magnetic charge is absent, regularity at the horizon allows only a trivial configuration for the axion field.
Thus, the background solution is given by the Reissner-Nordstr\"{o}m (RN) metric~\cite{Reissner:1916cle}, {\it i.e.}, 
\begin{align}
&g^{(0)}_{\mu\nu}dx^\mu dx^\nu=-f(r)dt^2+\frac{1}{f(r)}dr^2+r^2
(d\theta^2+\sin^2\theta\,d\varphi^2)~, \label{eq:g_bkg} \\
&A^{(0)}_\mu dx^\mu=-\frac{Q}{r}dt~, \label{eq:A_bkg} \\
&\phi^{(0)}=0~, \label{eq:phi_bkg}
\end{align}
where
\begin{align}
    f(r)&=1-\frac{2M}{r}+\frac{Q^2}{r^2}~,
\end{align}
with $M\,(>0)$ and $Q$ the BH mass and charge, respectively.
When $0<|Q|<M$, the function~$f(r)$ has two distinct positive roots,
\begin{align}
r_\pm=M\pm\sqrt{M^2-Q^2}~,
\end{align}
where $r_+$ and $r_-$ correspond to the event horizon and the Cauchy horizon, respectively.
We identify $r_+$ with $r_{\rm h}$ in Sec.~\ref{sec:formulation}.

\subsection{Master equations for perturbations}

Although the Chern-Simons term in Eq.~\eqref{eq:action_EMA} does not contribute to the background solution, it induces coupling among the degrees of freedom at the perturbation level.
We now linearize the field equations~\eqref{eq:Eeq}--\eqref{eq:KGeq} around the background solutions~\eqref{eq:g_bkg}--\eqref{eq:phi_bkg}, following Ref.~\cite{Boskovic:2018lkj}.
In particular, we introduce perturbations as
\begin{align}
&g_{\mu\nu}=g^{(0)}_{\mu\nu}+h_{\mu\nu}~, \\
&A_\mu=A^{(0)}_\mu+\delta A_{\mu}~, \\
&\phi=\phi^{(0)}+\delta\phi~.
\end{align}
Also, we expand the perturbation functions in a complete basis of spherical harmonics, and consider the frequency domain as defined in Eq.~\eqref{eq:Laplace} in the following, omitting the hat notation for brevity.

For the gravitational perturbation, in the Regge-Wheeler gauge~\cite{Regge:1957td},
\begin{align}
h_{\mu\nu}=\begin{pmatrix}
fH_0^{l}Y^{lm} & H_1^{l}Y^{lm} & h_0^{l}S_{\theta}^{lm} & h_0^{l}S_{\varphi}^{lm} \\
\ast & f^{-1}H_2^{l}Y^{lm} & h_1^{l}S_{\theta}^{lm} & h_1^{l}S_{\varphi}^{lm} \\
\ast & \ast & r^2 K^{l}Y^{lm} & 0 \\
\ast & \ast & \ast & r^2K^{l}\sin^2\theta Y^{lm}
\end{pmatrix}e^{-i\omega t}~,
\end{align}
where the asterisks represent symmetric components, $Y^{lm}=Y^{lm}(\theta,\varphi)$ are the spherical harmonics, and $(S_\theta^{lm},S_\varphi^{lm})=(Y_{,\varphi}^{lm}/\sin\theta,-\sin\theta Y_{,\theta}^{lm})$ are the axial vector harmonics.
For the EM field,
\begin{align}
\delta A_{\mu}=\begin{pmatrix}
u_{1}^l Y^{lm} \\
u_{2}^l Y^{lm} \\
u_{3}^l Y_b^{lm}
\end{pmatrix}e^{-i\omega t}
+\begin{pmatrix}
0 \\
0 \\
u_{4}^l S_b^{lm}
\end{pmatrix}e^{-i\omega t}~,
\end{align}
where $b=(\theta,\varphi)$, and $Y_b^{lm}=(Y_{,\theta}^{lm},Y_{,\varphi}^{lm})$ are the polar vector harmonics.
By use of the remaining gauge freedom associated with $U(1)$ symmetry, we impose $u_3^l=0$.
For the axion field,
\begin{align}
\delta\phi=\frac{\psi^l}{r}Y^{lm}e^{-i\omega t}~.
\end{align}
Here, the harmonic indices $l$ and $m$ are implicitly summed over.
It should be noted that we have omitted the index~$m$ for the perturbation variables as the evolution of perturbations does not depend on $m$ on a spherical background.
Also, in what follows, we shall omit the index~$l$ as modes with different $l$ evolve independently.

The perturbation variables can be classified as ``polar" and ``axial" based on their behavior under a parity transformation.
In particular, because the axion couples to the EM field through the Chern-Simons term, the axion perturbation only couples to the axial sector of the EM perturbation.
As a result, only the axial sector is influenced by the presence of the axion coupling.
Therefore, in the following, we focus only on the axial sector.
The polar sector remains unchanged from the Einstein-Maxwell case and has been comprehensively studied, for example, in Refs.~\cite{Chandrasekhar:1985kt,Burton:2020wnj}.

As a master variable for the axial sector, we introduce
\begin{align}
h^{\rm odd}=\frac{2}{(l-1)(l+2)}\left[r\left(r^2 \frac{d}{dr}\left(\frac{h_0}{r^2}\right)+i \omega h_1\right)+\frac{4Q}{r}u_4\right]~.
\end{align}
From $(t,\theta)$ and $(r,\theta)$ components of the first-order Einstein equation, we can get
\begin{align}
h_0&=\frac{r f}{2}\frac{d h^{\rm odd}}{dr}+\frac{f}{2}h^{\rm odd}~, \\
h_1&=-\frac{i\omega r}{2f}h^{\rm odd}~.
\end{align}
Thus, from $(r,\theta)$ component of the first-order Einstein equation and $\theta$ component of the first-order Maxwell equation, we obtain
\begin{align}
&\frac{d^2 h^{\rm odd}}{dr_\ast^2}+\left(\omega ^2-\frac{f \left(\left(l^2+l-3\right)r^2+Q^2\right)}{r^4}-\frac{3f^2}{r^2}\right)h^{\rm odd}+\frac{8 f Q}{r^3}u_4=0~, \\
&\frac{d^2 u_4}{dr_\ast^2}+\frac{f \left(l^2+l-2\right) Q}{2r^3}h^{\rm odd}+\left(\omega ^2-\frac{f \left(l (l+1)r^2+4 Q^2\right)}{r^4}\right)u_4+\frac{f g_{a\gamma\gamma}Q}{r^3}\psi=0~.
\end{align}
Here, we can diagonalize the potential term with respect to $(h^{\rm odd}, u_4)$ by performing a linear transformations from $(h^{\rm odd},u_4)$ to $(Z_+,Z_-)$ as
\begin{align}
h^{\rm odd}&=-2\frac{(3M-\lambda)Z_++(3M+\lambda)Z_-}{(l-1)(l+2)Q}~, \label{eq:hZ}\\
u_4&=Z_++Z_-~, \label{eq:uZ}
\end{align}
where $\lambda=\sqrt{9M^2+4Q^2(l-1)(l+2)}$.
Finally, combining the first-order Klein-Gordon equation, we obtain the master equation system for the axial sector as\footnote{The differences from the notation in Ref.~\cite{Boskovic:2018lkj} are as follows:
One of our master variables, $h^{\rm odd}$, corresponds to $\psi_{\rm RW}$ in Ref.~\cite{Boskovic:2018lkj}.
In addition, they denoted the axion-photon coupling as $k_{\rm a}$, which corresponds to $\ga/2$ in our notation.
Also, the sign convention for the electric charge~$Q$ is opposite.}
\begin{align}
&\frac{d^2 \psi}{dr_\ast^2}+(\omega^2-V_\psi)\psi+S_\psi Z_++S_\psi Z_-=0~, \label{eq:psi_pert} \\
&\frac{d^2 Z_+}{dr_\ast^2}+(\omega^2-V_+)Z_++S_+\psi=0~, \label{eq:Zp_pert} \\
&\frac{d^2 Z_-}{dr_\ast^2}+(\omega^2-V_-)Z_-+S_-\psi=0~, \label{eq:Zm_pert}
\end{align}
with
\begin{align}
V_\psi&=f(r)\left(\frac{l(l+1)}{r^2}+\frac{2M}{r^3}-\frac{2Q^2}{r^4}\right)~, \\
S_\psi&=f(r)\frac{l(l+1)g_{a\gamma\gamma}Q}{r^3}~, \label{eq:Spsi} \\
V_\pm&=f(r)\left(\frac{l(l+1)}{r^2}+\frac{-3M\pm\lambda}{r^3}+\frac{4Q^2}{r^4}\right)~, \\
S_\pm&=\pm f(r)\frac{g_{a\gamma\gamma}Q(3M\pm\lambda)}{2\lambda r^3}~, \label{eq:Spm}
\end{align}
where $r_\ast$ is the tortoise coordinate defined through $dr/dr_\ast=f(r)$.
It can be verified that this system of equations serves as a specific example of the general formulation discussed in Sec.~\ref{sec:formulation} for the case of $N=3$.
Notice that, in the limit~$g_{a\gamma\gamma}\to 0$, the three equations are decoupled.
Let us also comment on the limit~$Q\to 0$.
It should be noted that one cannot take $Q\to 0$ in Eq.~\eqref{eq:hZ}.
Nevertheless, the master equations~\eqref{eq:psi_pert}--\eqref{eq:Zm_pert} make sense in the limit~$Q\to 0$ and, as expected, they reduce to the standard Regge-Wheeler equations on a Schwarzschild background for fields with spin~0, 1, and 2, respectively.
In this case, $Z_+$ and $Z_-$ respectively correspond to purely EM and gravitational degrees of freedom.
In what follows, $\psi$, $Z_+$, and $Z_-$ are called axion-, EM-, and gravity-led modes, respectively.

\subsection{Quasinormal modes}

In this subsection, we compute the QNMs of BHs described by the coupled equation system~\eqref{eq:psi_pert}--\eqref{eq:Zm_pert}.
In particular, we examine whether resonant structure appears in the QNM frequency spectra.
Furthermore, we discuss the instability of the BH system for a large coupling suggested in Ref.~\cite{Boskovic:2018lkj}.

\subsubsection{Continued fraction method}\label{subsec:CF}

First, we describe the computation method used in this paper.
A representative calculation method is the direct integration method~\cite{Pani:2013pma}, which is widely applicable in many cases.
However, it is not suitable for extracting the mode functions that decay exponentially at large distances, making the computation of overtone modes particularly challenging.
Thus, we implement the continued fraction method developed by Leaver~\cite{Leaver:1985ax,Leaver:1990zz} for the EMA system.

From the definition discussed in Sec.~\ref{subsec:formulation_QNM}, the boundary conditions for QNMs are imposed as
\begin{align}
\Psi_\alpha\to\begin{cases}
e^{i\omega r_\ast}\sim r^{2iM\omega}e^{i\omega r}~, & r\to\infty~, \\
e^{-i\omega r_\ast}\sim (r-r_+)^{-\frac{i\omega r_+^2}{r_+-r_-}}~, \quad & r\to r_+~.
\end{cases}
\end{align}
Here, we denote $\vec{\Psi}=(\Psi_1,\Psi_2,\Psi_3)^{\rm t}=(\psi,Z_+,Z_-)^{\rm t}$.
It should be noted that Eqs.~\eqref{eq:psi_pert}--\eqref{eq:Zm_pert} take the form~$d^2\Psi_\alpha/dr_*^2+\omega^2\Psi_\alpha=0$ in the limit~$r_*\to\pm\infty$.
Following the boundary conditions, we firstly assume the ansatz for the master variables as
\begin{align}\label{eq:CF_ansatz}
\vec{\Psi}=r(r-r_-)^{-1+2iM\omega}e^{i\omega r}\left(\frac{r-r_+}{r-r_-}\right)^{-\frac{i\omega r_+^2}{r_+-r_-}}\sum_{j=0}^{\infty} 
\vec{a}_j
\left(\frac{r-r_+}{r-r_-}\right)^{j}~,
\end{align}
where $\vec{a}_j=(a_j^{(1)},a_j^{(2)},a_j^{(3)})^{\rm t}$ are constant vectors.
From the condition that $\sum_j\vec{a}_j$ should converge, the QNM frequencies can be determined.

Then, substituting the ansatz~\eqref{eq:CF_ansatz} into Eqs.~\eqref{eq:psi_pert}--\eqref{eq:Zm_pert}, we can obtain a finite-term recurrence relation for the series coefficients~$\vec{a}_j$.
In the present case, we obtain a five-term recurrence relation.
However, using the matrix-valued version of Gaussian elimination~\cite{Pani:2013pma,Nomura:2021efi}, it can be reduced to a three-term recurrence relation, allowing us to apply the matrix-valued version of Leaver's continued fraction method.
We summarize the details with explicit expression for the coefficients of the recurrence relation in App.~\ref{App:CF}.
As a result, $\vec{a}_j$ satisfies the recurrence relation given by
\begin{align}
    \bm{\alpha}_0\vec{a}_1+\bm{\beta}_0\vec{a}_0=0&~,     \label{eq:CFn0} \\
    \bm{\alpha}_j\vec{a}_{j+1}+\bm{\beta}_j\vec{a}_j+\bm{\gamma}_j\vec{a}_{j-1}=0& \quad (j\geq1)~. \label{eq:CFngt0}
\end{align}
Here, we introduce the ladder matrix~$\bm{R}_j^+$ such that
\begin{align}
    \vec{a}_{j+1}=\bm{R}_j^+\vec{a}_j~.
\end{align}
From Eq.~\eqref{eq:CFngt0}, we obtain
    \begin{align}
    \brb{\bra{\bm{\beta}_{j+1}+\bm{\alpha}_{j+1}\bm{R}_{j+1}^+}\bm{R}_j^{+}+\bm{\gamma}_{j+1}}\vec{a}_{j}=0~.
    \end{align}
Given that $\vec{a}_j$ is an arbitrary vector, we find the following recurrence relation of $\bm{R}_{j}^+$:
\begin{align}\label{eq:Rop}
    \bm{R}_j^+=-\left(\bm{\beta}_{j+1}+\bm{\alpha}_{j+1}\bm{R}_{j+1}^+\right)^{-1}\bm{\gamma}_{j+1}~.
\end{align}
Meanwhile, Eq.~\eqref{eq:CFn0} can be rewritten as
\begin{align}
    \bm{M}\vec{a}_0=0~ \quad {\rm with} \quad \bm{M}=\bm{\beta}_0+\bm{\alpha}_0\bm{R}_0^+~.
\end{align}
As the condition for the existence of nontrivial eigenvectors, we require
\begin{align}
    \det\bm{M}=0~. \label{detM=0}
\end{align}
Here, $\bm{R}_0^+$ in $\bm{M}$ is written in terms of $\bm{R}_1^+$ through Eq.~\eqref{eq:Rop}, and similarly, $\bm{R}_1^+$ is
written in terms of $\bm{R}_2^+$, and so on, and hence $\bm{M}$ is written in the form of a matrix-valued continued fraction.
For the values of $\omega$ that correspond to the QNMs, the continued fraction converges, and one can truncate it at some order~$N_{\rm max}$ by setting $\bm{R}_{N_{\rm max}}^+= \bm{I}$.
Now, Eq.~\eqref{detM=0} provides us with an algebraic equation for $\omega$, and the QNM frequencies can be obtained as its roots.
We take $N_{\rm max}$ to be sufficiently large for the solutions to converge.

\subsubsection{Numerical results}

Here, we show numerical results of the QNM frequencies of the BH in the EMA system.
In particular, we present results only for $l=2$, as it is the smallest $l$ where all the degrees of freedom are coupled and is also considered observationally important.
This system has three degrees of freedom, which correspond to purely gravitational, EM, and axion modes in the decoupling limit.
We refer to the QNMs that branch from each mode in the decoupling limit as gravity-led, EM-led, and axion-led modes for arbitrary parameter region.

\begin{figure*}[t]
\includegraphics[keepaspectratio, scale=0.8]{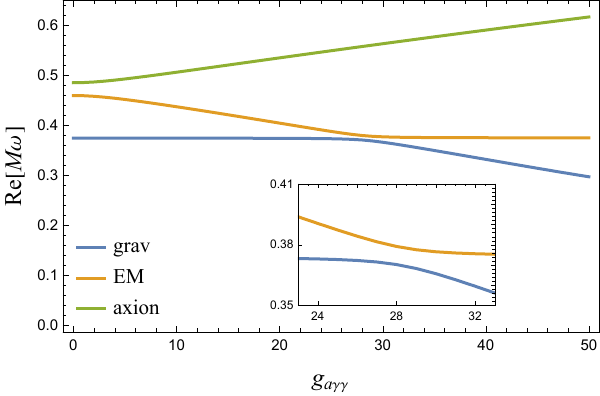}
\includegraphics[keepaspectratio, scale=0.8]{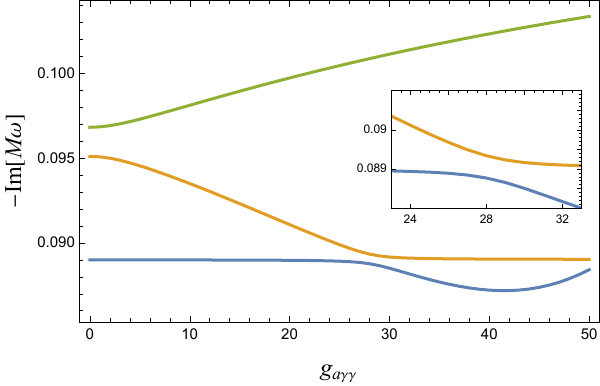}
\caption{
QNMs originating from the fundamental modes of the three degrees of freedom as functions of 
$\ga$ with $l=2$ and $Q/M=0.1$. The left panel shows the real part of the QNM frequencies, while the right panel shows the imaginary part. The blue, orange, and green lines correspond to the gravitational-, EM-, and axion-led modes, respectively. 
Around $\ga\simeq28$, avoided crossing between the gravity-led and EM-led
modes occurs.
}
\label{fig:QNMDoFQ01}
\end{figure*}

\begin{figure*}[t]
\includegraphics[keepaspectratio, scale=0.8]{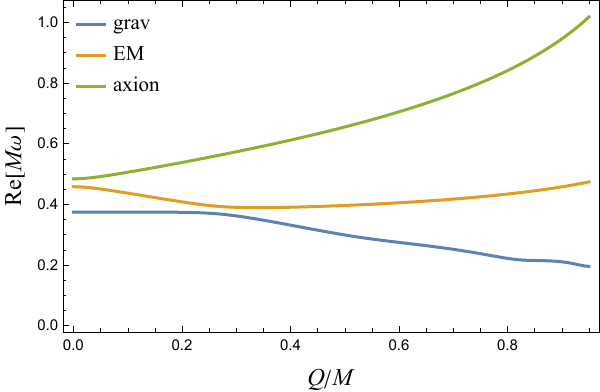}
\includegraphics[keepaspectratio, scale=0.8]{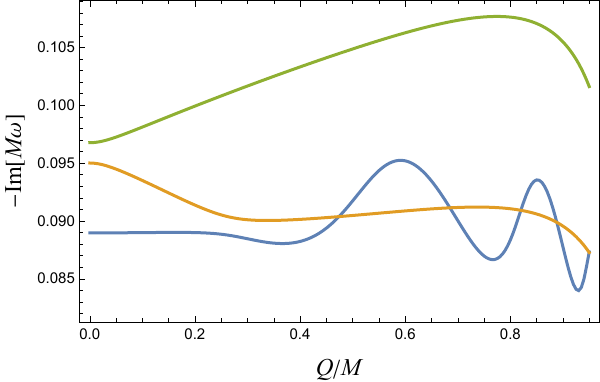}
\caption{
QNMs originating from the fundamental modes of the three degrees of freedom as functions of 
$Q/M$ with $l=2$ and $\ga=10$. The left panel shows the real part of the QNM frequencies, while the right panel shows the imaginary part. The meaning of each color is the same as in Fig.~\ref{fig:QNMDoFQ01}. $Q/M$ is varied from 0 to 0.95 in the plotted data.}
\label{fig:QNMDoFg10}
\end{figure*}

In Fig.~\ref{fig:QNMDoFQ01}, 
QNMs originating from the fundamental modes of the three degrees of freedom are shown as functions of the axion-photon coupling~$\ga$ with $Q/M=0.1$.
The blue, orange, and green lines correspond to the gravity-led, EM-led, and axion-led modes, respectively.
In particular, in the limit~$\ga\to0$, they reduce to those in the RN BH case.
Focusing on the gravity-led and EM-led modes, as $\ga$ increases from 0, both ${\rm Re}[\omega]$ and $-{\rm Im}[\omega]$ of the former remain almost unchanged for $\ga\lesssim28$, while those of the latter decrease.
Then, around $\ga\simeq28$, the gravity-led and EM-led QNMs repel each other as avoided crossing.
For $\ga\gtrsim28$, at least immediately after the avoided crossing, one mode appears to follow along the branch of the other.

In addition, Fig.~\ref{fig:QNMDoFg10} shows the results for varying $Q$ with fixed $\ga=10$.
In this case as well, although somewhat less clearly, avoided crossing between the gravity-led and EM-led QNMs is observed around $Q/M\simeq0.26$ in both the real and imaginary parts.
For smaller $\ga$, a larger value of $Q$ is required for avoided crossing to occur.
In particular, when $\ga=0$ (the RN BH case), the gravity-led mode and EM-led mode decouple, and therefore, avoided crossing does not occur.

\begin{figure}[t]
\includegraphics[keepaspectratio, scale=0.8]{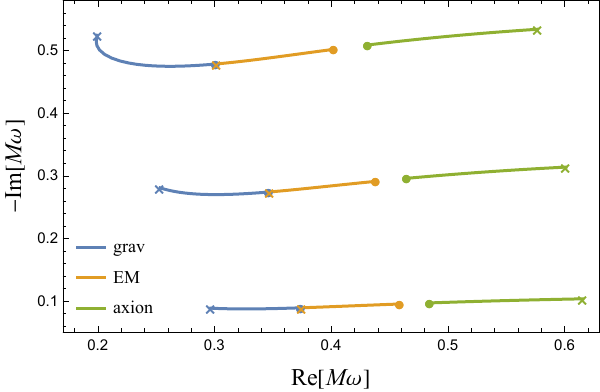}
\includegraphics[keepaspectratio, scale=0.8]{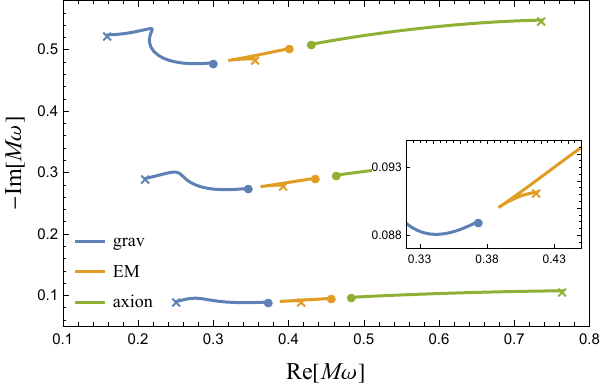}
\caption{QNM frequencies for $l=2$ in the complex plane, including up to the second overtone, {\it i.e.}, $n=0$, $1$, and $2$ (from bottom to up). The blue, orange, and green lines correspond to the gravity-led, EM-led, and axion-led modes, respectively. In the left panel, the axion-photon coupling~$\ga$ is varied from 0 to 50 with fixed $Q/M=0.1$ in the plotted data. In the right panel, the BH charge~$Q/M$ is varied from 0 to 0.7 with fixed $\ga=10$. In both panels, the starting points are indicated by circles and the endpoints by crosses.}
\label{fig:QNMovertone}
\end{figure}

In Fig.~\ref{fig:QNMovertone}, we also show trajectories of the QNM frequencies in the complex plane, including up to the second overtone, {\it i.e.}, $n=0$, 1, and 2.
Each color corresponds to the same mode as those figures.
Note that the data shown correspond to the parameter ranges where $\ga$ varies from 0 to 50 with fixed $Q/M=0.1$ in the left panel and $Q/M$ varies from 0 to 0.7 with fixed $\ga=10$ in the right panel.
In this region, the overall behavior is similar for all QNMs with $n=0$, 1, and 2.
Interestingly, in the case where $\ga$ is varied (left panel), the QNM frequencies of each degree of freedom belonging to the same overtone number in the decoupling limit appear to 
move along a single curve in the complex plane.
In particular, the gravity-led QNMs move in such a way that they are pushed away by the EM-led QNMs (exactly like billiards), as a result of the avoided crossing.
This avoided crossing occurs at nearly the same parameter values across all overtones as far as we have investigated.
We will return to a more detailed discussion of the QNM spectra in a later section.
It should also be noted that, differently from the Kerr case, the avoided crossing occurs between longest-lived modes originating from the fundamental modes of different degrees of freedom.
This is obviously a phenomenon specific to the coupled system, and it is necessary to verify whether resonant excitation can occur.

\subsubsection{Instability}

\begin{figure}[t]
\includegraphics[keepaspectratio, scale=0.8]{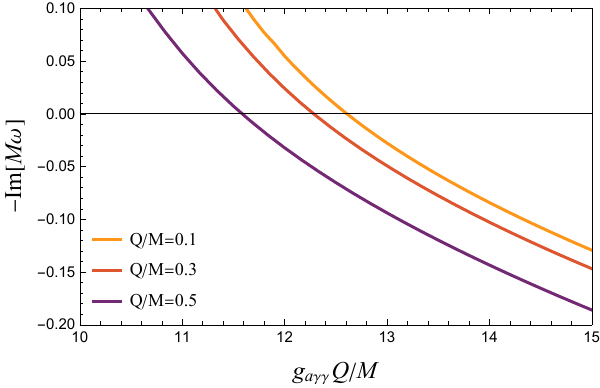}
\caption{Imaginary part of the non-oscillatory mode frequencies as a function of $\ga Q/M$ with $l=2$. The results for $Q/M=0.1$, 0.3, and 0.5 with varying $\ga$ are shown.
}
\label{fig:QNMinst}
\end{figure}

In the regime where $g_{a\gamma\gamma}Q/M$ is relatively large, the real part of the gravity-led QNM frequency tends to decrease beyond the point where the avoided crossing occurs.
In this regime, the convergence of the continued fraction method with respect to the truncation order becomes increasingly poor, especially for higher overtones.
Moreover, as investigated in Ref.~\cite{Boskovic:2018lkj}, the system exhibits non-oscillatory modes whose 
frequency is purely imaginary.
The imaginary parts of such modes are shown in Fig.~\ref{fig:QNMinst}.
As demonstrated, the imaginary part becomes positive once $\ga Q/M$ exceeds a certain value; for example, when $Q/M=0.1$, this occurs for $\ga\gtrsim126$.

This indicates that, in the regime where the coupling is relatively large, not only does the numerical computation of QNMs become challenging, but the system also becomes unstable against linear perturbations.
It should be noted that this type of instability is not a general feature of coupled systems but is specific to the EMA system~\cite{Molina:2010fb}.
Since the study of such instabilities is not the primary focus of this paper, we leave related investigation for future works.
In the following, we restrict our analysis to the region where the system remains stable and the coupling is rather small.

\subsection{Excitation factors}
As seen in the previous subsection, since the QNM frequencies exhibit the avoided crossing, resonant excitation of QNMs is expected in this system.
In this subsection, we present the explicit expression of the excitation factor and verify it through numerical calculation and analytical interpretation.

Based on the formulation given in Sec.~\ref{subsec:formulation_EF}, we provide the definition of the excitation factor for the EMA system.
First, we define the ``in-mode" function whose asymptotic behavior is given by
\begin{align}
\vec{\Psi}^{({\rm in},\alpha)}=
\begin{pmatrix}
\psi^{({\rm in},\alpha)} \\
Z_+^{({\rm in},\alpha)} \\
Z_-^{({\rm in},\alpha)}
\end{pmatrix}\to
\begin{cases}
e^{-i\omega r_\ast}\vec{\delta}_\alpha~, &r_\ast\to-\infty~, \\
e^{-i\omega r_\ast}\vec{A}^{({\rm in},\alpha)}+e^{+i\omega r_\ast}\vec{A}^{({\rm out},\alpha)}~, \quad &r_\ast\to+\infty~,
\end{cases}
\end{align}
where $\vec{\delta}_\alpha=(\delta_{\alpha1},\delta_{\alpha2},\delta_{\alpha3})^{{\rm t}}$, $\vec{A}^{({\rm in},\alpha)}=(A_\psi^{({\rm in},\alpha)},A_{Z_+}^{({\rm in},\alpha)},A_{Z_-}^{({\rm in},\alpha)})^{\rm t}$ and $\vec{A}^{({\rm out},\alpha)}=(A_\psi^{({\rm out},\alpha)},A_{Z_+}^{({\rm out},\alpha)},A_{Z_-}^{({\rm out},\alpha)})^{\rm t}$.
Then, the coefficient matrix for the in-going wave $\bm{A}^{({\rm in})}$ is defined as
\begin{align}
\bm{A}^{({\rm in})}=\begin{pmatrix}
A_\psi^{({\rm in},1)} & A_\psi^{({\rm in},2)} & A_\psi^{({\rm in},3)} \\
A_{Z_+}^{({\rm in},1)} & A_{Z_+}^{({\rm in},2)} & A_{Z_+}^{({\rm in},3)} \\
A_{Z_-}^{({\rm in},1)} & A_{Z_-}^{({\rm in},2)} & A_{Z_-}^{({\rm in},3)}
\end{pmatrix}~.
\end{align}
Similarly, the coefficient matrix for the out-going wave $\bm{A}^{({\rm out})}$ is defined by replacing ``in" with ``out" in the above equation.
Using the above quantities, the excitation factor is defined as
\begin{align}
B_{\alpha n}=\frac{i\det\bm{A}^{({\rm out})}}{(2i\omega_{\alpha n})^3}\left.\left(\frac{d}{d\omega}\det\bm{A}^{({\rm in})}\right)^{-1}\right|_{\omega=\omega_{\alpha n}}~.
\end{align}
Note that the determinant of the coefficient matrices can be calculated through the relation with the Wronskian matrix;
\begin{align}
\det\bm{A}^{({\rm in})}&=(2i\omega)^{-3}\det \bm{W}[\vec{\Psi}^{({\rm in},1)},\vec{\Psi}^{({\rm in},2)},\vec{\Psi}^{({\rm in},3)},\vec{\Psi}^{({\rm up},1)},\vec{\Psi}^{({\rm up},2)},\vec{\Psi}^{({\rm up},3)}]~, \\
\det\bm{A}^{({\rm out})}&=-(2i\omega)^{-3}\det \bm{W}[\vec{\Psi}^{({\rm in},1)},\vec{\Psi}^{({\rm in},2)},\vec{\Psi}^{({\rm in},3)},\vec{\Psi}^{({\rm down},1)},\vec{\Psi}^{({\rm down},2)},\vec{\Psi}^{({\rm down},3)}]~,
\end{align}
where the ``up-mode" and ``down-mode" satisfy $\vec{\Psi}^{({\rm up},\alpha)}\to e^{+i\omega r_\ast}\vec{\delta}_\alpha$ and $\vec{\Psi}^{({\rm down},\alpha)}\to e^{-i\omega r_\ast}\vec{\delta}_\alpha$ for $r_\ast\to\infty$, respectively.

\subsubsection{Numerical results}

To obtain the excitation factor, it is necessary to compute the coefficient matrix, which in turn requires solving the homogeneous solutions under each boundary condition.
First, we briefly describe the numerical calculation method.
As discussed in the previous subsection, the resonant excitation is expected among the 
QNMs originating from the fundamental modes of the gravity-led and EM-led degrees of freedom. 
Therefore, in the following, we focus on
such longest-lived modes and compute the mode function using the direct integration method~\cite{Pani:2013pma} based on higher-order expansions of $\Psi_\alpha$ around both the horizon and infinity.
Then, we adopt the ansatz 
\begin{align}
    \Psi_\alpha=(r-r_+)^{-\frac{i\omega r_+^2}{r_+-r_-}}\sum_{j=0}^{\infty}(\Psi_\alpha^{\rm H})^{(j)}(r-r_+)^j~,
\end{align}
around the horizon and
\begin{align}
    \Psi_\alpha=r^{\pm2iM\omega}e^{\pm i\omega r}\sum_{j=0}^{\infty}(\Psi_\alpha^{\rm I})^{(j)}r^{-j}~,
\end{align}
at infinity.
Note that, in practical computation, we truncate the summation at a sufficiently large $j$.
Solving the perturbation equations order by order, we find the coefficients~$(\Psi_\alpha^{\rm H/I})^{(j)}$ as functions of $(\Psi_\alpha^{\rm H/I})^{(0)}$. 
For each mode function to be determined, we specify $(\Psi_\alpha^{\rm H/I})^{(0)}$, impose the boundary conditions from the above equations, and numerically solve the perturbation equations.

\begin{figure*}[t]
\includegraphics[keepaspectratio, scale=0.8]{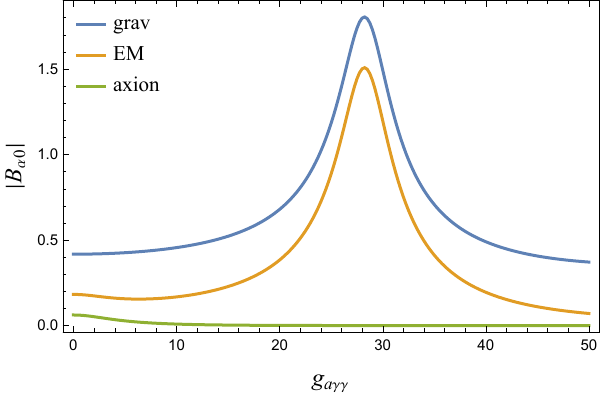}
\includegraphics[keepaspectratio, scale=0.8]{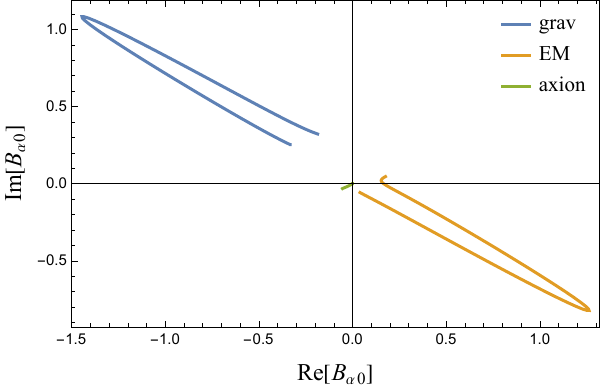}
\caption{Excitation factors for the QNMs originating from the fundamental modes of the three degrees of freedom with $l=2$ and $Q/M=0.1$. The left panel shows the absolute value of excitation factors as functions of the axion-photon coupling~$\ga$. The right panel shows the excitation factors in the complex plane when $\ga$ varies from 0 to 50. The blue, orange, and green lines correspond to the excitation factors for the gravity-led ($B_{g0}$), EM-led ($B_{e0}$), and axion-led QNMs ($B_{a0}$), respectively. Around $\ga\simeq28$, where the avoided crossing between the gravity-led mode and EM-led mode occurs in the left panel of Fig.~\ref{fig:QNMDoFQ01}, amplification of the excitation factors is observed.}
\label{fig:EF1}
\end{figure*}

Here, we present numerical results of the excitation factors.
Figure~\ref{fig:EF1} shows the excitation factors $B_{\alpha 0}$ for the fundamental QNMs of each degree of freedom with $l=2$ and $Q/M=0.1$.
In the left panel, the absolute values of the excitation factors $|B_{\alpha 0}|$ 
are plotted as functions of the axion-photon coupling $\ga$.
As expected from the behavior of the QNM spectra in the left panel of Fig.~\ref{fig:QNMDoFQ01}, the amplification of the excitation factors corresponding to the gravity-led mode and EM-led mode can be observed around $\ga\simeq28$, where the avoided crossing of 
the QNMs occurs.

The right panel of Fig.~\ref{fig:EF1} also presents the excitation factors in the complex plane when $\ga$ varies from 0 to 50.
In the case of Kerr QNMs, as shown in Ref.~\cite{Motohashi:2024fwt}, excitation factors almost trace the lemniscate of Bernoulli in the complex plane.
In particular, near the exceptional point where avoided crossing occurs between different overtones, the associated excitation factors also exhibit approximately point-symmetric amplification in the complex plane.
On the other hand, in the EMA case, where spectral repulsion occurs between QNMs originating from different degrees of freedom, the trajectory of the excitation factor does not resemble a lemniscate, unlike the Kerr case.
However, it also exhibits a point-symmetric amplification in the complex plane, which verifies the resonant excitation of the QNMs.

\begin{figure}[t]
\includegraphics[keepaspectratio, scale=0.55]{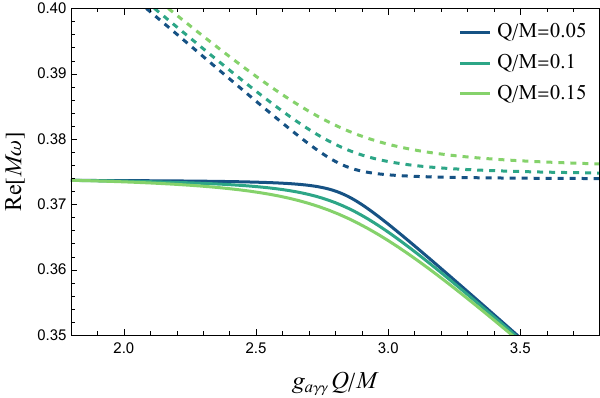}
\includegraphics[keepaspectratio, scale=0.55]{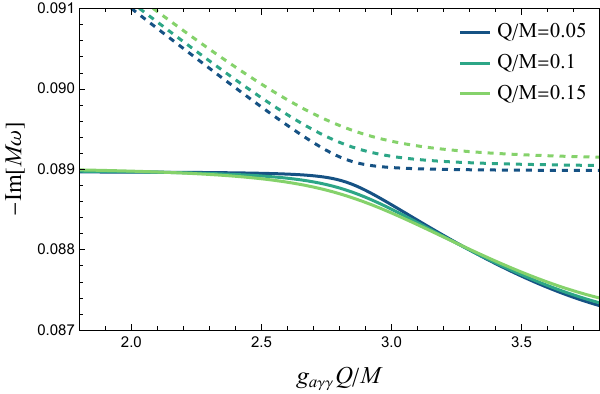}
\includegraphics[keepaspectratio, scale=0.55]{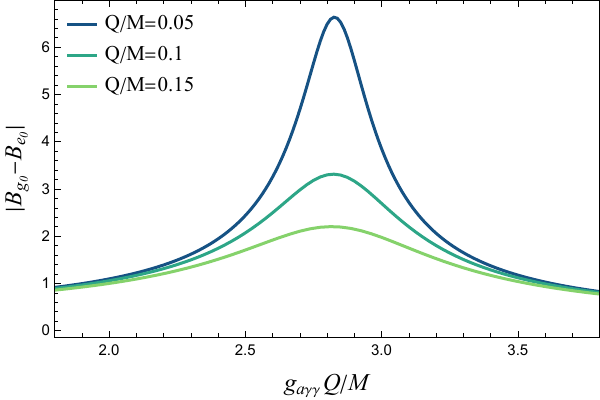}
\caption{Real (left) and imaginary (middle) parts of the QNM frequencies originating from the fundamental modes of the gravity-led (solid) and EM-led (dashed) 
modes with $l=2$ and different value of $Q/M$. The right panel shows the absolute value of the difference between their corresponding excitation factors.}
\label{fig:EF2}
\end{figure}

\begin{figure*}[t]
\includegraphics[keepaspectratio, scale=0.8]{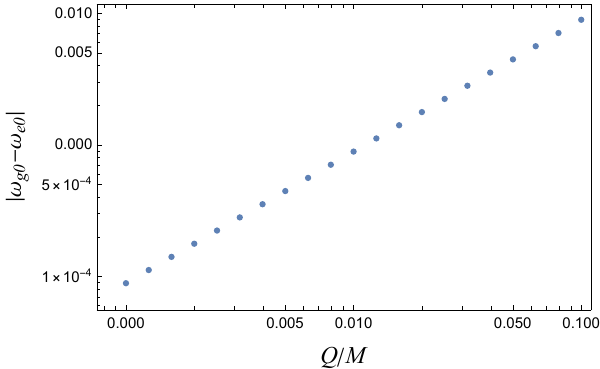}
\includegraphics[keepaspectratio, scale=0.8]{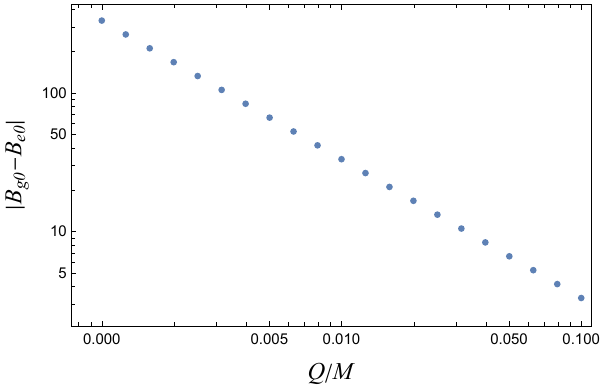}
\caption{Absolute values of the difference between QNM frequencies of gravity-led and EM-led modes (left) and their corresponding excitation factors (right) as functions of $Q/M$. The value of $\ga Q/M$ is fixed at 2.8284, where the avoided crossing occurs.}
\label{fig:Qdep}
\end{figure*}

In addition, we show the change in resonance behavior as the BH charge~$Q$ varies in Fig.~\ref{fig:EF2}.
The real parts (left) and imaginary parts (middle) of the gravity-led and the EM-led fundamental QNM frequencies are shown.
These are denoted as $\omega_{g0}$ and $\omega_{e0}$, and the absolute value of the difference between their corresponding excitation factors~$|B_{g0}-B_{e0}|$ is shown in the right panel.
From Eqs.~\eqref{eq:psi_pert}--\eqref{eq:Zm_pert}, as far as $Q/M\ll1$, the dominant contribution to the coupling between the degrees of freedom appears through the combination of $\ga Q/M$.
Therefore, by plotting them as functions of $\ga Q/M$, we can confirm that both the avoided crossing and excitation occur at approximately the same value of $\ga Q/M$.
On the other hand, the width of the spectral repulsion and excitation factors depend on $Q$ itself.
The smaller the value of $Q$, the narrower the width of the spectral repulsion.
As shown in the right panel, this leads to a greater amplification of the excitation factor at the peak.
As shown in Fig.~\ref{fig:Qdep}, we find that, at the point where avoided crossing occurs, there are approximate relations~$|\omega_{g0}-\omega_{e0}|\propto Q$ and $|B_{g0}-B_{e0}|\propto Q^{-1}$ with fixed value of $\ga Q/M$.
Note that resonance at small $Q/M$ requires a large axion-photon coupling~$\ga$.

\subsubsection{
Key feature of the resonant excitation}

\begin{figure}[t]
\includegraphics[keepaspectratio, scale=0.55]{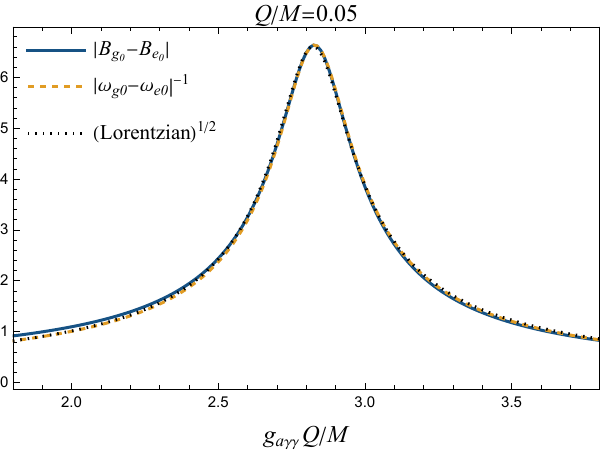}
\includegraphics[keepaspectratio, scale=0.55]{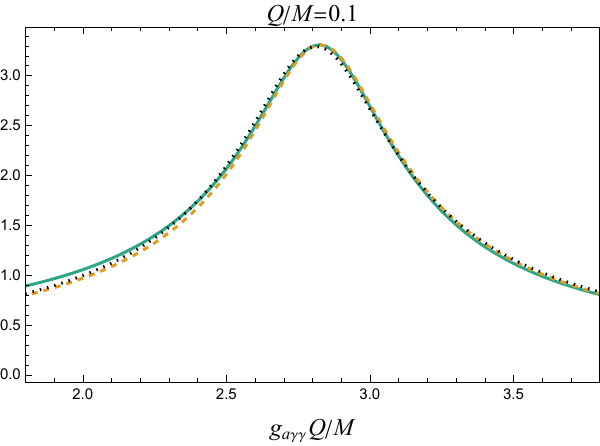}
\includegraphics[keepaspectratio, scale=0.55]{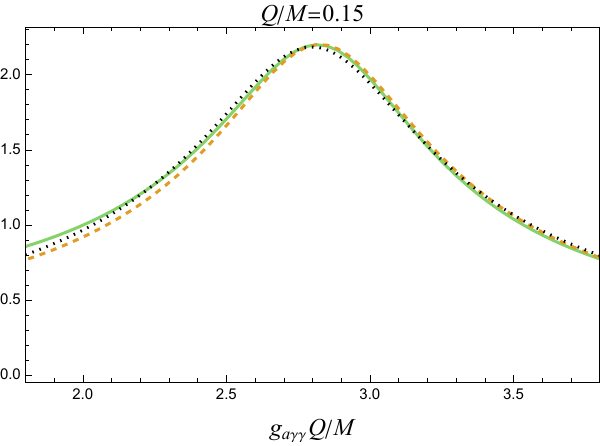}
\caption{Absolute values of the difference between the excitation factors (solid) and the inverse of the difference between QNM frequencies $|\omega_{g0}-\omega_{e0}|^{-1}$ (orange dashed), with the latter normalized to have the same maximum, along with a fit by a square-root of Lorentzian function (black dotted). The values of $Q/M$ are 0.05, 0.1 and 0.15 from left to right.
}
\label{fig:EFfit}
\end{figure}

Finally, we discuss the resonance behaviors of QNM spectra and the excitation factors.
Although QNMs are difficult to treat analytically, their features can be captured by the following simple model.
As proposed in Ref.~\cite{Motohashi:2024fwt}, the theory of QNMs can be well understood in terms of quantum mechanics for non-Hermitian systems.
Around the avoided crossing, since two eigenvalues are close to degenerate, it is appropriate to consider a two-level system.
Here, we denote eigenvalues of EM-led and gravity-led QNMs in the absence of the direct coupling between them as ${\cal E}_1$ and ${\cal E}_2$, respectively.
As a model to describe the eigenvalues $\omega_\pm$ in the presence of the coupling, we employ~\cite{Motohashi:2024fwt}
\begin{align}\label{eq:model}
    \omega_\pm^2={\cal E}_c\pm\sqrt{{\cal E}_d^2+\Delta^2}~,
\end{align}
where
\begin{align}
    {\cal E}_c=\frac{1}{2}({\cal E}_1+{\cal E}_2)~, \quad {\cal E}_d=\frac{1}{2}({\cal E}_1-{\cal E}_2)~.
\end{align}

Focusing on the case where $\ga$ is varied with fixed $Q/M$,\footnote{In the case where $Q/M$ is varied with fixed $\ga$, the behavior of the QNM spectra becomes more complicated as shown in the right panel of Fig.~\ref{fig:QNMovertone}. The interpretation of this is left for future work.} to reproduce the spectral behavior shown in the left panel of Fig.~\ref{fig:QNMovertone}, we simply assume\footnote{Compared to the model in Ref.~\cite{Motohashi:2024fwt}, the phase of these quantities are set to be aligned.}
\begin{align}\label{eq:model_para}
    {\cal E}_1=p e^{i\vartheta}~, \quad  {\cal E}_2=\omega_0^2 e^{i\vartheta}~, \quad \Delta=q e^{i\vartheta}~,
\end{align}
where $\omega_0$ and $\vartheta$ are real constants, while $p$ and $q$ are varying real parameters.
In the case of the EMA system, even if the coupling to the gravity-led mode is neglected, the EM-led mode couples to the axion-led mode through Eq.~\eqref{eq:Spsi}.
Therefore, the parameter $p$, which characterizes the variation of ${\cal E}_1$, is proportional to $\ga Q$.
Substituting Eq.~\eqref{eq:model_para} into Eq.~\eqref{eq:model}, we obtain
\begin{align}\label{eq:model2}
    \omega_\pm^2=e^{i\vartheta}\left[\frac{1}{2}\left(p+\omega_0^2\right)\pm\sqrt{\frac{1}{4}\left(p-\omega_0^2\right)^2+q^2}\right]~.
\end{align}
It is straightforward to verify that in the limit~$p/\omega_0^2\to \infty$ with $q/\omega_0^2$ fixed, 
$\omega_-^2\to{\cal E}_1$ and $\omega_+^2\to{\cal E}_2$.
On the other hand, in the limit~$p/\omega_0^2\to -\infty$ with $q/\omega_0^2$ fixed,
$\omega_-^2\to{\cal E}_2$ and $\omega_+^2\to{\cal E}_1$.
As expected, Eq.~\eqref{eq:model2} exhibits avoided crossing in both the real and imaginary parts, and in the complex plane, one mode that was varying appears to push out the other, which had remained constant.

In Ref.~\cite{Motohashi:2024fwt}, it was demonstrated that the relation between excitation factors and eigenvalues is given by
\begin{align}\label{eq:EFmodel}
    |B_+-B_-|\propto|\omega_+-\omega_-|^{-1}~.
\end{align}
In the case of the Kerr QNM model, the eigenvalues follow a hyperbolic trajectory near the avoided crossing, and the excitation factors take the form of a quarter-power Lorentzian function in a nontrivial manner. 
However, in the case of the present model, the difference of the eigenvalues is given by
\begin{align}\label{eq:diffomega2}
    |\omega_+^2-\omega_-^2|=2\sqrt{\frac{1}{4}\left(p-\omega_0^2\right)^2+q^2}~.
\end{align}
Therefore, from Eqs.~\eqref{eq:EFmodel} and \eqref{eq:diffomega2}, the difference of the excitation factors is given by the square-root of a Lorentzian function.
From the behavior~$|\omega_{g0}-\omega_{e0}|\propto Q$ at the point corresponding to the avoided crossing ($p\simeq\omega_0^2$) shown in Fig.~\ref{fig:Qdep}, it can be inferred that $q$ is proportional to $Q$. This is due to the indirect coupling between gravity-led and EM-led modes through the axion-led mode.

We show in Fig.~\ref{fig:EFfit} the difference between excitation factors corresponding to the gravity-led and EM-led QNMs along with the inverse of the difference between QNM frequencies~$|\omega_{g0}-\omega_{e0}|^{-1}$.
We also include a fit using a square-root of Lorentzian function, $f(\ga Q/M)=f_0/\{(\ga Q/M-x_0)^2+\gamma^2\}^{1/2}$, where $f_0$, $x_0$, and $\gamma$ are positive constants.
They show excellent agreement, especially when $Q$ is small and the amplification is sharp.
This result confirms the validity of the model and the resonant excitation.

\section{Conclusions}\label{sec:conc}
In this paper, we have developed a general formalism for excitation factors of BH QNMs in systems where multiple degrees of freedom are coupled. 
Based on the Green's matrix method, we have extended the single-degree-of-freedom framework to define excitation factors in a way that is independent of the choice of the basis of variables.
This formulation enables the investigation of resonant amplification in QNMs of coupled systems.

We have applied this formulation to BHs in the Einstein-Maxwell-axion system as a concrete example.
On the 
Reissner-Nordstr{\"o}m BH background, the three degrees of freedom---gravitational, EM, and axion fields---are coupled through the perturbation equations for the axial sector.
We have computed the QNMs originating from each degree of freedom, including higher overtones, using the continued fraction method.
When the axion-photon coupling~$\ga$ or the BH charge~$Q$ is varied, we have found avoided crossing between the gravity-led and EM-led QNMs in both the real and imaginary parts.
This is a repulsion phenomenon between QNMs originating from different degrees of freedom, in contrast to the repulsion between different overtones of a single degree of freedom observed in the Kerr case in General Relativity.
Moreover, this avoided crossing occurs almost simultaneously across overtones, 
and importantly, it also takes place between 
longest-lived modes originating from fundamental modes of the gravity-led and EM-led modes.
The repulsion in both the real and imaginary parts differs from avoided crossing between eigenvalues of a single degree of freedom in general non-Hermitian systems, for which it has been known that typically either the real parts or imaginary parts cross~\cite{PhysRevE.61.929}.

We have verified the resonance behavior by computing the excitation factors formulated in this work.
Specifically, we have found that the excitation factors associated with the relevant QNMs are amplified at the parameter values where the avoided crossing occurs.
Moreover, those corresponding to the gravity-led and EM-led QNMs exhibit a point-symmetric amplification in the complex plane, reminiscent of the Kerr case.
However, the trajectories of the 
QNMs and excitation factors in the present system differ from those in the Kerr case.
Motivated by the framework of the non-Hermitian quantum mechanics, we have employed a simple 
model of the QNM trajectories, which provides an excellent fit to the behavior of the excitation factors near the resonance.
These results confirm the existence of a new type of resonance phenomenon arising from the coupling 
among multiple degrees of freedom, distinct from the 
resonance among overtones observed in the Kerr case.

Such resonance phenomena can offer a novel perspective for probing gravitational theories or new particles through ringdown gravitational waves.
Our formulation is applicable to systems with coupled degrees of freedom in general and provides a framework for identifying resonances associated with avoided crossings in other theoretical contexts as well.
In systems with more degrees of freedom, even more intricate resonance structures may emerge.
In this work, our findings are based on frequency-domain analysis.
To establish a clearer connection with observable quantities and to explore phenomenological implications of resonance, it is essential to perform time-domain waveform calculations in more realistic setups.
The extraction of resonance structure by data analysis techniques remains an important direction for future work.


\acknowledgments{
We would like to thank Richard Brito for useful correspondence.
This work was supported in part by JSPS (Japan Society for the Promotion of Science) KAKENHI Grant Nos.~JP23KJ1214, JP25KJ0067, and JP25K17397 (T.T.), No.~JP22K03639 (H.M.), and No.~JP23K13101 (K.T.).
}


\appendix

\section{Green's matrix}\label{App:Green}
In this appendix, we summarize the Green's matrix method~\cite{Reid:1931} applied in Sec.~\ref{subsec:formulation_EF}.
Especially, we derive the construction of the Green's matrix shown in Eq.~\eqref{eq:Green}, following Ref.~\cite{Sisman:2009mk}.
Let us consider an $N$-component vector~$\vec{y}(x)$ governed by a second-order differential equation system in the general
form, and then construct the solution under boundary conditions at two points.
Thus, the differential equation system is given by
\begin{align}\label{eq:second-diff}
\hat{\bm{M}}\vec{y}(x)=\vec{h}(x)~,
\end{align}
where
\begin{align}
\hat{\bm{M}}=\begin{pmatrix}
\hat{M}_{1} & V_{12}(x) & \cdots & V_{1N}(x) \\
V_{21}(x) & \hat{M}_{2} & \cdots & V_{2N}(x) \\
\vdots & \vdots & \ddots & \vdots \\
V_{N1}(x) & V_{N2}(x) & \cdots & \hat{M}_{N}
\end{pmatrix}~, 
\end{align}
with
\begin{align}
\hat{M}_{\alpha}&=\left[\frac{d}{dx}\left(p_\alpha(x)\frac{d}{dx}\right)+q_\alpha(x)\right]~.
\end{align}
The boundary conditions are imposed at $x=a$ and $x=b$, with
\begin{align}\label{eq:BC}
\vec{y}^{\,\prime}(a)=\bm{\alpha}\vec{y}(a)\quad{\rm and} \quad \vec{y}^{\,\prime}(b)=\bm{\beta}\vec{y}(b)~,
\end{align}
respectively, where a prime denotes the derivative with respect to $x$.
The Green's matrix~$\bm{G}(x,\tilde{x})$ of the differential operator~$\hat{\bm{M}}$ is defined with the formal solution
\begin{align}\label{eq:G_app}
    \vec{y}(x)=\int_a^b d\tilde{x}\ \bm{G}(x,\tilde{x})\vec{h}(\tilde{x})~.
\end{align}

First, defining a $2N$-vector
\begin{align}
\vec{z}(x)=\begin{pmatrix}
\vec{y} \\ \vec{y}^{\,\prime}
\end{pmatrix}~, 
\end{align}
the second-order differential equation system~\eqref{eq:second-diff} is rewritten as a first-order differential equation system
\begin{align}\label{eq:first_diff}
\vec{z}^{\,\prime}(x)=\bm{A}(x)\vec{z}(x)+\vec{f}(x)~.
\end{align}
Here, introducing the selection matrices
\begin{align}
\bm{U}^{\rm L}_{N\times 2N}=\begin{pmatrix} \bm{I} & \bm{O} \end{pmatrix}~, \quad \bm{L}^{\rm D}_{2N\times N}=\begin{pmatrix} \bm{O} \\ \bm{I} \end{pmatrix}~,
\end{align}
$\vec{y}$ and $\vec{f}$ are expressed as
\begin{align}
&\vec{y}(x)=\bm{U}^{\rm L}_{N\times 2N}\vec{z}(x)~, \label{eq:yz} \\
&\vec{f}(x)=\bm{L}^{\rm D}_{2N\times N}\bm{P}^{-1}(x)\vec{h}(x)~, \label{eq:fh}
\end{align}
where
\begin{align}
\bm{P}(x)={\rm diag}\left[p_1(x),p_2(x),\cdots,p_N(x)\right]~.
\end{align}
For the construction of the Green's matrix, the explicit expression of $\bm{A}$ in Eq.~\eqref{eq:first_diff} is not needed (but can be found in the Ref.~\cite{Sisman:2009mk}).

Next, we assume a specific solution of the form
\begin{align}\label{eq:z_app}
\vec{z}(x)=\bm{W}(x)\vec{g}(x)~,
\end{align}
where $\bm{W}$ is a $2N\times 2N$ matrix constructed from $2N$ homogeneous solutions which satisfy $\vec{z}^{\,\prime}=\bm{A}\vec{z}$, {\it i.e.},
\begin{align}
\bm{W}'(x)=\bm{A}(x)\bm{W}(x)~. \label{defW}
\end{align}
Substituting Eq.~\eqref{eq:z_app} into the differential equation system with the source~\eqref{eq:first_diff}, a formal solution can be obtained as
\begin{align}
\vec{g}(x)=\int_a^x d\tilde{x}\ \bm{W}^{-1}(\tilde{x})\vec{f}(\tilde{x})+\vec{c}~,
\end{align}
where $\vec{c}$ is an integration constant vector.
Therefore, as a solution of the first differential equation system, we obtain
\begin{align}\label{eq:formal}
\vec{z}(x)=\bm{W}(x)\int_a^x d\tilde{x}\ \bm{W}^{-1}(\tilde{x})\vec{f}(\tilde{x})+\bm{W}(x)\vec{c}~.
\end{align}

Now, we rewrite the boundary conditions~\eqref{eq:BC} as
\begin{align}
\bm{B}_a\vec{z}(a)+\bm{B}_b\vec{z}(b)=\vec{0}~,
\end{align}
where
\begin{align}
\bm{B}_a=\begin{pmatrix} 
-\bm{\alpha} & \bm{I} \\
\bm{O} & \bm{O}
\end{pmatrix}~, \quad
\bm{B}_b=\begin{pmatrix} 
\bm{O} & \bm{O} \\
-\bm{\beta} & \bm{I}
\end{pmatrix}~.
\end{align}
Note that the matrices~$\bm{B}_a$ and $\bm{B}_b$ are fixed only up to an overall factor.
Nevertheless, as shown below, this ambiguity does not affect the final result of the Green's matrix.
From these boundary conditions, $\vec{c}$ is fixed as
\begin{align}
\vec{c}=-\bm{D}^{-1}\bm{B}_b\bm{W}(b)\int_a^b d\tilde{x}\ \bm{W}^{-1}(\tilde{x})\vec{f}(\tilde{x})~,
\end{align}
where $\bm{D}$ is defined by
\begin{align}
\bm{D}=\bm{B}_a\bm{W}(a)+\bm{B}_b\bm{W}(b)~.
\end{align}
Substituting it into the general solution~\eqref{eq:formal}, we obtain
\begin{align}
\vec{z}(x)=\bm{W}(x)\int_a^x d\tilde{x}\ \bm{W}^{-1}(\tilde{x})\vec{f}(\tilde{x})-\bm{W}(x)\bm{D}^{-1}\bm{B}_b\bm{W}(b)\int_a^b d\tilde{x}\ \bm{W}^{-1}(\tilde{x})\vec{f}(\tilde{x})~.
\end{align}
After some algebra, it can be rewritten as
\begin{align}
\vec{z}(x)=\bm{W}(x)\bm{D}^{-1}\bm{B}_a\bm{W}(a)\int_a^x d\tilde{x}\ \bm{W}^{-1}(\tilde{x})\vec{f}(\tilde{x})-\bm{W}(x)\bm{D}^{-1}\bm{B}_b\bm{W}(b)\int_x^b d\tilde{x}\ \bm{W}^{-1}(\tilde{x})\vec{f}(\tilde{x})~.
\end{align}
As a result, the Green's matrix for the first-order differential equation system defined through
\begin{align}
\vec{z}=\int_a^b d\tilde{x}\ \bm{G}_1(x,\tilde{x})\vec{f}(\tilde{x})~,
\end{align}
can be expressed as
\begin{align}
\bm{G}_1(x,\tilde{x})=\begin{cases}\label{eq:G1}
-\bm{W}(x)\bm{D}^{-1}\bm{B}_b\bm{W}(b)\bm{W}^{-1}(\tilde{x})~, \quad &x<\tilde{x}~, \\
\bm{W}(x)\bm{D}^{-1}\bm{B}_a\bm{W}(a)\bm{W}^{-1}(\tilde{x})~, \quad &x>\tilde{x}~.
\end{cases}
\end{align}

Let us now go back to the second-order system~\eqref{eq:second-diff} and define the basis of homogeneous solutions where the expression of the Green's matrix is simplified.
We consider $2N$ linearly independent homogeneous solutions of Eq.~\eqref{eq:second-diff}, of which $N$ satisfy the boundary condition~$\vec{y}^{\,\prime}(a)=\bm{\alpha}\vec{y}(a)$ and the remaining $N$ satisfy $\vec{y}^{\,\prime}(b)=\bm{\beta}\vec{y}(b)$, {\it i.e.},
\begin{align}
&\hat{\bm{M}}\vec{u}^{\,\alpha}(x)=\vec{0}~, \quad \vec{u}^{\,\alpha\prime}(a)=\bm{\alpha}\vec{u}^{\,\alpha}(a)~, \\
&\hat{\bm{M}}\vec{v}^{\,\beta}(x)=\vec{0}~, \quad \vec{v}^{\,\beta\prime}(b)=\bm{\beta}\vec{v}^{\,\beta}(b)~,
\end{align}
where $\alpha,\beta=1,2,\cdots,N$.
Defining
\begin{align}
&\bm{U}=\begin{pmatrix}
\vec{u}^{\,1},\vec{u}^{\,2},\cdots,\vec{u}^{\,N}
\end{pmatrix}~, \\
&\bm{V}=\begin{pmatrix}
\vec{v}^{\,1},\vec{v}^{\,2},\cdots,\vec{v}^{\,N}
\end{pmatrix}~,
\end{align}
the Wronskian matrix can be constructed as
\begin{align}
\bm{W}=\begin{pmatrix}
\bm{U} &\bm{V} \\
\bm{U}' & \bm{V}'
\end{pmatrix}~.
\end{align}
Note that each column is constructed out of a homogeneous solution to Eq.~\eqref{eq:second-diff} and its derivative, and therefore this $\bm{W}$ can be identified as $\bm{W}$ defined in Eq.~\eqref{defW} for the first-order system.
Written in this form, we have
\begin{align}\label{eq:DBW}
\bm{D}^{-1}\bm{B}_a\bm{W}(a)=\begin{pmatrix}
\bm{O} & \bm{O} \\
\bm{O} & \bm{I}
\end{pmatrix}~, \quad 
\bm{D}^{-1}\bm{B}_b\bm{W}(b)=\begin{pmatrix}
\bm{I} & \bm{O} \\
\bm{O} & \bm{O}
\end{pmatrix}~,
\end{align}
which simplify the expression in Eq.~\eqref{eq:G1}.

From the above, we can obtain the Green's matrix for the second-order differential equation system defined by Eq.~\eqref{eq:G_app}.
Substituting Eq.~\eqref{eq:DBW} into Eq.~\eqref{eq:G1} and using Eqs.~\eqref{eq:yz} and~\eqref{eq:fh}, we can construct the Green's matrix as
\begin{align}
\bm{G}(x,\tilde{x})=\begin{cases}
-\bm{U}^{\rm L}_{N\times 2N}\bm{W}_a(x)\bm{W}^{-1}(\tilde{x})\bm{L}^{\rm D}_{2N\times N}\bm{P}^{-1}(\tilde{x})~, \quad &x<\tilde{x}~, \\
\bm{U}^{\rm L}_{N\times 2N}\bm{W}_b(x)\bm{W}^{-1}(\tilde{x})\bm{L}^{\rm D}_{2N\times N}\bm{P}^{-1}(\tilde{x})~, \quad &x>\tilde{x}~, 
\end{cases}
\end{align}
where
\begin{align}
\bm{W}_a(x)=\bm{W}(x)\begin{pmatrix}
\bm{I} & \bm{O} \\
\bm{O} & \bm{O}
\end{pmatrix}~, \quad
\bm{W}_b(x)=\bm{W}(x)\begin{pmatrix}
\bm{O} & \bm{O} \\
\bm{O} & \bm{I}
\end{pmatrix}~.
\end{align}


\section{Coefficients of the recurrence relation}\label{App:CF}

In this appendix, we present the explicit expression for the recurrence relation coefficients in the continued fraction method used in Sec.~\ref{subsec:CF} to compute the QNMs of a BH in the EMA system.
The original recurrence relation obtained from Eqs.~\eqref{eq:psi_pert}--\eqref{eq:Zm_pert} is a five-term recurrence relation:
\begin{align}
\bm{\alpha}_{0}\vec{a}_{1}+\bm{\beta}_{0}\vec{a}_{0}&=0~,  \\
\bm{\alpha}_{1}\vec{a}_{2}+\bm{\beta}_{1}\vec{a}_{1}+\bm{\gamma}_{1}\vec{a}_{0}&=0~, \label{5rec1} \\
\bm{\alpha}_{2}\vec{a}_{3}+\bm{\beta}_{2}\vec{a}_{2}+\bm{\gamma}_{2}\vec{a}_{1}+\bm{\delta}_{2}\vec{a}_{0}&=0~, \label{5rec2} \\
\bm{\alpha}_{j}\vec{a}_{j+1}+\bm{\beta}_{j}\vec{a}_{j}+\bm{\gamma}_{j}\vec{a}_{j-1}+\bm{\delta}_{j}\vec{a}_{j-2}+\bm{\epsilon}_{j}\vec{a}_{j-3}&=0 \quad  (j\geq 3)~. \label{5rec3}
\end{align}
The coefficient matrices are given as follows:
\begin{align}
\bm{\alpha}_j=\begin{pmatrix}
\alpha_j^\psi & 0 & 0 \\
0 & \alpha_j^{Z} & 0 \\
0 & 0 & \alpha_j^{Z}
\end{pmatrix}~,
\end{align}

\begin{align}
&\alpha_j^\psi=(j+1) r_+ \left(M (j+1)+r_+\left(-j+i r_+ \omega-1\right)\right)~, \\
&\alpha_j^{Z}=-r_+\alpha_j^\psi~,
\end{align}

\begin{align}
\bm{\beta}_j=\begin{pmatrix}
\beta_j^\psi & l (l+1) g_{a\gamma \gamma} Q \left(M-r_+\right) & l (l+1) g_{a\gamma \gamma} Q \left(M-r_+\right) \\
-\frac{g_{a\gamma\gamma}Q r_+ (3 M+\lambda)\left(M-r_+\right)}{2 \lambda } & \beta_j^{Z_+} & 0 \\
\frac{g_{a\gamma \gamma}Q r_+ (3 M-\lambda )\left(M-r_+\right)}{2 \lambda } & 0 & \beta_j^{Z_-}
\end{pmatrix}~,
\end{align}

\begin{align}
\beta_j^\psi=&\:M r_+ \left(-l (l+1)+j^2+2 i j r_+\omega -2 j+2 i r_+ \omega-1\right) \notag \\
&+r_+^2 \left(l(l+1)+j^2-5 i j r_+ \omega +2 j-4r_+^2 \omega ^2-3 i r_+ \omega+1\right)-2 M^2 j^2~, \\
\beta_j^{Z_\pm}=&\:r_+ \left\{M \left(\pm\lambda +r_+\left(l (l+1)-4 j^2+2j-12\right)-2 i r_+^2 \omega\right)\right. \notag \\
&\left.+r_+ \left(\mp\lambda -r_+(l (l+1)+2 j-5)+i (4 j+3) r_+^2\omega +4 r_+^3 \omega^2\right)+M^2 \left(4j^2+7\right)\right\}~,
\end{align}

\begin{align}
\bm{\gamma}_j=\begin{pmatrix}
\gamma_j^\psi & -l (l+1) g_{a\gamma \gamma} Q \left(M-r_+\right) & -l (l+1) g_{a\gamma \gamma} Q \left(M-r_+\right) \\
\frac{g_{a\gamma\gamma}M Q  \left(M-r_+\right)(3 M+\lambda)}{ \lambda } & \gamma_j^{Z_+} & 0 \\
-\frac{g_{a\gamma \gamma}M Q \left(M-r_+\right)(3 M-\lambda )}{ \lambda } & 0 & \gamma_j^{Z_-}
\end{pmatrix}~,
\end{align}

\begin{align}
\gamma_j^\psi=&\:2 M^2 \left(l (l+1)+2 j^2-6 i j r_+ \omega -2 j+2 i r_+\omega +1\right) \notag \\ 
&+M r_+ \left(-3 l (l+1)-5 j^2+20 i jr_+ \omega +6 j+12 r_+^2 \omega ^2-8 i r_+ \omega-3\right)  \notag \\
&+r_+^2 \left(l (l+1)+j^2-5 i j r_+ \omega -2j-4 r_+^2 \omega ^2+3 i r_+ \omega +1\right)~, \\
\gamma_j^{Z_\pm}=&-2 M^2 \left(\pm\lambda +r_+ (2 l (l+1)+4 j+15)-2 i (4 j-1)r_+^2 \omega \right) \notag \\
&+M r_+ \left(\pm2 \lambda +r_+ \left(6l (l+1)+6 j^2-2 j+37\right)-4 i (8 j-3) r_+^2 \omega-20 r_+^3 \omega ^2\right) \notag \\
&+r_+^3 \left(-2 l (l+1)-2j^2+10 i j r_+ \omega +2 j+8 r_+^2 \omega ^2-5 i r_+\omega -13\right)+M^3 \left(-4 j^2+8 j+6\right)~,
\end{align}

\begin{align}
\bm{\delta}_j=\begin{pmatrix}
\delta_j^\psi & 0 & 0 \\
-\frac{g_{a\gamma\gamma}Q  \left(M-r_+\right)(2M-r_+)(3 M+\lambda)}{ 2\lambda } & \delta_j^{Z_+} & 0 \\
\frac{g_{a\gamma\gamma}Q  \left(M-r_+\right)(2M-r_+)(3 M-\lambda)}{ 2\lambda } & 0 & \delta_j^{Z_-}
\end{pmatrix}~,
\end{align}

\begin{align}
\delta_j^\psi=&\:\left(2 M-r_+\right) \left\{4 i M^2 (j-1) \omega -M\left(j^2+j \left(-2+4 i r_+ \omega \right)+4 r_+^2\omega ^2-4 i r_+ \omega +1\right)\right. \notag \\
&\left.+(j-1) r_+ \left(j-ir_+ \omega -1\right)\right\}~, \\
\delta_j^{Z_\pm}=&\:\left(2 M-r_+\right) \left\{M^2 \left(2 l (l+1)+4 j^2-16 ij r_+ \omega -12 j+20 i r_+ \omega +5\right)\right. \notag \\
&\left.+M\left(\lambda -r_+ \left(3 l (l+1)+4 j^2-14j+2\right)+8 i (3 j-4) r_+^2 \omega +16 r_+^3 \omega^2\right)\right. \notag \\
&\left.-r_+ \left(\pm\lambda +r_+ (-l (l+1)+2 j+3)+i (4j-7) r_+^2 \omega +4 r_+^3 \omega ^2\right)\right\}~,
\end{align}

\begin{align}
\bm{\epsilon}_j=\begin{pmatrix}
0 & 0 & 0 \\
0 & \epsilon_j^{Z} & 0 \\
0 & 0 & \epsilon_j^{Z}
\end{pmatrix}~,
\end{align}

\begin{align}
\epsilon_j^{Z}=\left(r_+-2 M\right){}^2 \left(-M
   \left(j^2+j \left(-4+4 i r_+
   \omega \right)+4 r_+^2 \omega
   ^2-8 i r_+ \omega +4\right)+4 i
   (j-2) M^2 \omega +(j-2) r_+
   \left(j-i r_+ \omega
   -2\right)\right)~.
\end{align}

The five-term recurrence relation can be reduced to a three-term relation by applying the matrix-valued version of Gaussian elimination as we demonstrate below.
First, the term with $\vec{a}_0$ in Eq.~\eqref{5rec2} can be removed by use of Eq.~\eqref{5rec1} as
    \begin{align}
    \bm{\alpha}_{2}\vec{a}_{3}+\bm{\beta}'_{2}\vec{a}_{2}+\bm{\gamma}'_{2}\vec{a}_{1}=0~, \label{5rec2'}
    \end{align}
with
    \begin{align}
    \bm{\beta}'_2=\bm{\beta}_2-\bm{\delta}_2\bm{\gamma}_1^{-1}\bm{\alpha}_1~, \quad
    \bm{\gamma}'_2=\bm{\gamma}_2-\bm{\delta}_2\bm{\gamma}_1^{-1}\bm{\beta}_1~.
    \end{align}
Then, for $j=3$ of Eq.~\eqref{5rec3}, the last two terms with $\vec{a}_0$ and $\vec{a}_1$ can be removed by use of Eqs.~\eqref{5rec1} and \eqref{5rec2'} as
    \begin{align}
    \bm{\alpha}_{3}\vec{a}_{4}+\bm{\beta}'_{3}\vec{a}_{3}+\bm{\gamma}'_{3}\vec{a}_{2}=0~, 
    \end{align}
with
    \begin{align}
    \bm{\beta}'_3=\bm{\beta}_3-\bm{\delta}'_3\bm{\gamma}_2^{\prime-1}\bm{\alpha}_2~, \quad
    \bm{\gamma}'_3=\bm{\gamma}_3-\bm{\epsilon}_3\bm{\gamma}_1^{-1}\bm{\alpha}_1-\bm{\delta}'_3\bm{\gamma}_2^{\prime-1}\bm{\beta}'_2~, \quad
    \bm{\delta}'_3=\bm{\delta}_3-\bm{\epsilon}_3\bm{\gamma}_1^{-1}\bm{\beta}_1~.
    \end{align}
Likewise, one can remove the last two terms in Eq.~\eqref{5rec3} for arbitrary $j$'s.
Written explicitly, for $j\ge 4$, we obtain
    \begin{align}
    \bm{\alpha}_{j}\vec{a}_{j+1}+\bm{\beta}'_{j}\vec{a}_{j}+\bm{\gamma}'_{j}\vec{a}_{j-1}=0~, 
    \end{align}
with
    \begin{align}
    \bm{\beta}'_j=\bm{\beta}_j-\bm{\delta}'_j\bm{\gamma}_{j-1}^{\prime-1}\bm{\alpha}_{j-1}~, \quad
    \bm{\gamma}'_j=\bm{\gamma}_j-\bm{\epsilon}_j\bm{\gamma}_{j-2}^{\prime-1}\bm{\alpha}_{j-2}-\bm{\delta}'_j\bm{\gamma}_{j-1}^{\prime-1}\bm{\beta}'_{j-1}~, \quad
    \bm{\delta}'_j=\bm{\delta}_j-\bm{\epsilon}_j\bm{\gamma}_{j-2}^{\prime-1}\bm{\beta}'_{j-2}~.
    \end{align}
In Eq.~\eqref{eq:CFngt0} of the main text, $\bm{\beta}'_j$ and $\bm{\gamma}'_j$ defined above are respectively denoted as $\bm{\beta}_j$ and $\bm{\gamma}_j$ for simplicity.


\bibliographystyle{mybibstyle}
\bibliography{bib}

\end{document}